\documentclass[10pt,journal]{IEEEtran}
\usepackage{amsmath,amsfonts}
\usepackage{algorithmic}
\usepackage{algorithm}
\usepackage{array}
\usepackage{xcolor}
\usepackage{textcomp}
\usepackage{stfloats}
\usepackage{subfigure}
\usepackage{amsmath}
\usepackage{url}
\usepackage{multicol}
\usepackage{verbatim}
\usepackage{graphicx}
\usepackage{cite}
\begin{document}

\title{Addressing the Curse of Scenario and Task Generalization in AI-6G: A Multi-Modal Paradigm}

\author{Tianyu Jiao, Zhuoran Xiao, Yin Xu, Chenhui Ye, Yihang Huang, Zhiyong Chen, Liyu Cai, Jiang Chang,\\ Dazhi He, Yunfeng Guan, Guangyi Liu, and Wenjun Zhang,~\IEEEmembership{Fellow,~IEEE}
\thanks{Manuscript received xxx; revised xxx; accepted xxx. This work was supported in part by the National Key Research and Development Program of China under Grant 2024YFE0200600.}
\thanks{Tianyu Jiao (\textit{Student Member, IEEE}) is with the Cooperative Medianet Innovation Center, Shanghai Jiao Tong University, Shanghai 200240, China, and also with the Nokia Bell Labs, Shanghai 201206, China (e-mail: jiaotianyu@sjtu.edu.cn).}
\thanks{Zhuoran Xiao (\textit{Member, IEEE}), Chenhui Ye (\textit{Member, IEEE}), Yihang Huang (\textit{Member, IEEE}), Liyu Cai (\textit{Member, IEEE}), and Jiang Chang are with the Nokia Bell Labs, Shanghai 201206, China (e-mail: \{zhuoran.xiao, chenhui.a.ye, yihang.huang, liyu.cai, jiang.chang\}@nokia-sbell.com).}
\thanks{Yin Xu (\textit{Senior Member, IEEE}), Zhiyong Chen (\textit{Senior Member, IEEE}), Dazhi He (\textit{Senior Member, IEEE}), Yunfeng Guan, and Wenjun Zhang (\textit{Fellow, IEEE}) are with the Cooperative Medianet Innovation Center, Shanghai Jiao Tong University, Shanghai 200240, China (e-mail: \{xuyin, zhiyongchen, hedazhi, yfguan69, zhangwenjun\}@sjtu.edu.cn).}
\thanks{Guangyi Liu (\textit{Member, IEEE}) is with the China Mobile Research Institute, Beijing 100053, China (e-mail: liuguangyi@chinamobile.com).}
\thanks{The corresponding author is Yin Xu.}
}



\maketitle

\begin{abstract}
Existing works on machine learning (ML)-empowered wireless communication primarily focus on monolithic scenarios and single tasks. However, with the blooming growth of communication task classes coupled with various task requirements in future 6G systems, this working pattern is obviously unsustainable. Therefore, identifying a groundbreaking paradigm that enables a universal model to solve multiple tasks in the physical layer within diverse scenarios is crucial for future system evolution.
This paper aims to fundamentally address the curse of ML model generalization across diverse scenarios and tasks by unleashing multi-modal feature integration capabilities in future systems. 
Given the universality of electromagnetic propagation theory, the communication process is determined by the scattering environment, which can be more comprehensively characterized by cross-modal perception, thus providing sufficient information for all communication tasks across varied environments. This fact motivates us to propose a transformative two-stage multi-modal pre-training and downstream task adaptation paradigm.
In the pre-training stage, we introduce a multi-modal two-tower model and a corresponding contrastive learning method to integrate the explicit description of the scattering environment and implicit channel state information (CSI) into a universal representation, which encapsulates rich high-level knowledge and can be leveraged for all downstream tasks in different scenarios. Additionally, we present two specially designed model structures to enhance the interaction of communication modalities. In the second stage, based on the frozen pre-trained model, we propose a direct method and a pluggable method for flexible and low-cost task adaptation. Experimental results demonstrate that our proposed approach significantly outperforms benchmarks in both task performance and tuning parameter size for exemplary sub-tasks in unseen scenarios.
\end{abstract}


\section{Introduction}
\subsection{Motivation}
\IEEEPARstart{M}{achine} learning (ML) techniques are envisioned to be extensively utilized in 6G communication systems. However, due to changes in scattering environments, existing studies primarily focus on specialized models tailored to particular tasks, which are supervised-trained using simple data from a fixed scenario, such as a small area served by a specific base station (BS) \cite{r2}. Consequently, these specific models are scarcely transferable to other scenarios and tasks. 
Considering the difficulty of data collection in actual scenarios, especially high-quality training data, and the fact that the actual propagation environment is not static, these specific models need to be updated frequently. All these factors pose critical challenges to system flexibility and model computation, storage, and transmission overheads. In future 6G systems, functionality will be richer, and the propagation environment will be more complex. An exponential increase in the number of specialized models can be foreseen, which is evidently unsustainable.

If ubiquitous user equipment (UE) and BS could pre-store a universal model agnostic to specific scenarios and tasks, they would only need to make slight adjustments based on this universal model with minimal data and low computational costs to perform multiple tasks across various scenarios. Moreover, the universal model holds rich knowledge from different scenarios, which can enhance task performance in new, unseen environments.
Although environment transfer learning and multi-task learning methods have been attempted in a few studies, they still rely on fixed supervised training, limiting further generality and usability. Besides, updating the entire model is barely effective in reducing overall costs. Therefore, it is imperative to devise a more universal paradigm that can flexibly and cost-effectively generalize across different areas, BSs, and tasks while achieving comparable or even superior performance compared to specific models.

With the emergence and adaptation of integrated sensing and communication (ISAC), visual-assisted communication, speech command recognition, and the global positioning system (GPS) in 6G communication, the wireless system is entering a new era with significant performance potential leveraging multi-modal capabilities \cite{r0}. Firstly, the multi-modal model can fuse information from different modalities and capture common features, thereby improving performance and robustness. For example, in a noisy transmission channel, the accuracy of selecting the optimal beamforming codeword can be enhanced by combining visual information. Secondly, when only single-modal data is available, the network module for this modality can be utilized individually, greatly improving system flexibility. For instance, a beamforming task can be performed using position or channel data. Most importantly, the multi-modal model can generalize across its modalities, enabling it to handle unseen inputs and perform multiple tasks, thus enhancing user experience and quality of service (QoS).

Multi-modal universal models in the image and text formats, exemplified by contrastive language-image pre-training (CLIP) \cite{r1}, have demonstrated robust cross-modal semantic comprehension and generalization capabilities. Through self-supervised contrastive learning, related image-text features are drawn close while unrelated features are pushed far apart in the embedding space, thus learning rich task-agnostic universal representations. By understanding intrinsic semantic relationships between different modalities rather than merely achieving better fitting, multi-modal contrastive learning surpasses traditional end-to-end supervised learning. It is worth noting that the modalities for the 6G communication system are much richer than those in the computer science (CS) community, indicating much greater potential for overall performance \cite{r0}.

Despite the diversity in areas, BSs, and tasks, electromagnetic propagation theory serves as a consistent foundation determining the transmission process \cite{r2}. All potential modalities in wireless systems essentially encapsulate propagation path information. For example, channel state information (CSI) encompasses all the information of paths, serving as a basic channel modality for the wireless system. The explicit position and distance information of the BS, UE, and area map inherently contain extractable path parameters, such as angle, delay, and power, analogous to ray-tracing techniques, making them a typical environment modality in practical scenarios. Therefore, a 6G-oriented multi-modal universal model can be built by adequately comparing different communication modalities and extracting latent environment and propagation path information in a uniform embedding feature space, thus achieving generalization across areas, BSs, and tasks.

Leveraging the frozen multi-modal universal model, zero-shot learning (ZSL) and task-oriented fine-tuning (TOFT) methods can be designed to flexibly and cost-effectively perform multiple communication tasks as well as environment and position sensing tasks in various unseen areas and BSs. Specifically, the ZSL method can directly compare the similarity of unseen multi-modal data without tuning parameters. The TOFT method plugs a lightweight scenario- and task-specific network after the frozen universal model to transform task-agnostic universal representations into specific task objectives.

\subsection{Related Work}
\textit{1) Multi-Modal Models for 6G:} In 6G systems, intelligent communication and sensing rely on designing and deploying wireless multi-modal AI models \cite{r2}. Training a task-agnostic multi-modal telecom model using diverse telecom data equips it to handle various modalities such as camera, LiDAR, radar, GPS, and wireless channel data. Subsequent fine-tuning enables the model to perform a series of downstream tasks, including localization, beamforming, power allocation, and handover, even in unseen network scenarios. This method eliminates the need for dedicated AI models for each task \cite{r3}. Robustness to multi-modal problems could be improved by training global models that perform well on average across all tasks. This requires some distributed training, i.e., several local models, each specifically trained for one task \cite{rr1}\cite{rr2}\cite{rr3}. Currently, multi-modal communication data are solely utilized as multiple inputs for single-task supervised training \cite{r4}. Only some concepts are provisioned for upcoming wireless multi-modal models, lacking practical implementation in model design and task performance enhancement.

\textit{2) Bidirectional Mapping Relationship Between Physical Environment and Wireless Channel Modalities:} There is a bidirectional mapping relationship between environment semantics in an area, such as trees, roads, and buildings, and channel parameter semantics, such as angle of departure (AOD), angle of arrival (AOA), and the number of paths \cite{r5}. Based on the layouts of buildings and transmitters, neural networks (NNs) can be used to fit the path loss of the area, which is considered a pseudo ray-tracing process \cite{r6}. RGB environment images can be utilized to perform multiple wireless tasks such as vision-aided beam codebook design \cite{r7}, beam and blockage prediction \cite{r9}, and multi-user matching and resource allocation \cite{r10}. However, obtaining images requires deploying multiple cameras on BSs, UEs, and roadsides and capturing data from various directions, significantly increasing system complexity. On the other hand, CSI can provide AoD and time of arrival (ToA) information for each path. By combining this with the reflector positions in the area map and utilizing a ray reflection model to infer all propagation paths, the intersection of these paths indicates the UE position \cite{r11}. Furthermore, by applying reflection path extraction and clustering algorithms to analyze a series of radio signals received during movement, the surrounding environment can be reconstructed to a certain extent \cite{r12}. In summary, propagation path semantic information can be revealed from both the environment and channel modalities.

\textit{3) Environment Transfer and Multi-Tasking:} The transfer learning and multi-task learning methods in ML have been separately utilized to achieve environment adaptation and perform multiple tasks. A transfer learning framework named DDA-Net was proposed to achieve environment transfer for the CSI feedback task \cite{r13}. A meta-learning scheme was introduced to learn environment-independent features, aiding environment adaptation for the positioning task \cite{r14}. Multi-task learning was employed to perform two simple classification tasks: localization and identification recognition \cite{r15}, or to design a shared encoder with multiple task-specific decoders for the multi-scenario CSI feedback task \cite{r16}. To our knowledge, no study has yet simultaneously achieved generalization across areas, BSs, and tasks.

\textit{4) Contrastive Learning for CSI:} Contrastive learning enables the learning of discriminative and generalizable representations without labels. Considering the scarcity of labeled CSI data, contrastive learning is employed to learn CSI representations by contrasting positive and negative CSI samples through pre-training. Subsequently, human activity recognition (HAR) tasks \cite{r17}\cite{r18}\cite{r19} and positioning tasks \cite{r20} can be performed by fine-tuning a small NN appended to the pre-trained encoder using labeled data, or directly searching for similar fingerprints in the embedding space. However, these methods only investigate a single modality without considering the semantic correlations between different modalities.

\subsection{Contributions of This Work}
Motivated by the modal diversity of future 6G communication systems, the universality of electromagnetic propagation theory, and the success of image-text multi-modal models, we propose a novel multi-modal paradigm, which comprises a universal model capable of handling multi-modal communication data and direct ZSL and pluggable TOFT methods that can perform multiple tasks across unseen scenarios at low costs.

It needs to be emphasized that building a 6G-oriented multi-modal universal model is non-trivial because multi-modal data varies in precision and dimensions, implicit universal compressed representations should be physically interpretable, and task objectives have different formulations, data structures, and model optimizing strategies. Additionally, effective methods that are cost-efficient and data-friendly for adapting the frozen universal model to unseen environments need to be specially designed by exploiting specialized communication properties. Our proposed paradigm bridges the physical environment and wireless channel modalities, significantly supporting the development of future large-scale communication intelligence.

The main contributions are summarized as follows:
\begin{itemize}
\item We analyze the equivalence of propagation path information between the physical environment and wireless channel modalities, enabling cross-modal understanding and generalization. Based on this, we propose a 6G-oriented paradigm for multi-modal pre-training and downstream task adaptation across various unseen scenarios.
\item We devise a multi-modal universal model that aligns the physical environment and wireless channel modalities to achieve cross-scenario capability and extract task-agnostic universal representations through pre-training. 
An environment perception neural network (EPNN) and a channel feature extraction neural network (CFENN) are designed to extract features from their respective modalities. Both of them use a modality-shared neural network (MSNN) to promote modality interaction and fusion. The extracted representations of both modalities are aligned in the same embedding space through contrastive learning. To ensure the integrity of path information for these features, a channel reconstruction neural network (CRNN), together with a reconstruction loss, is incorporated.
\item In the downstream task adaptation stage, leveraging low-dimensional universal modality representations, we propose a ZSL method for direct classification and sensing and a TOFT method that plugs in lightweight specific NNs to perform multiple tasks. Experimental results show that our paradigm achieves superior accuracy with few or no tuning parameters in four exemplary downstream tasks across various unseen scenarios, even on unseen datasets.
\end{itemize}

The remainder of this paper is organized as follows. Section II introduces the considered system model and channel model, and analyzes why the physical environment and the wireless channel modalities can be contrasted. Section III presents the design of the proposed multi-modal pre-training and downstream task adaptation paradigm. Section IV describes our experimental setup, evaluates the pre-trained universal model, and compares the performance of our paradigm with that of multiple benchmarks. Finally, Section V draws the conclusion.

\section{System Model and Feasibility Analysis}
In this section, first, we introduce the differences between the proposed universal paradigm and traditional scenario- and task-specific methods in actual system design. Then, we describe the channel model of the massive multiple-input multiple-output (MIMO) system. Finally, we analyze the equivalence of propagation path information between the physical environment and wireless channel modalities.

\begin{figure*}[t]
\centering
\includegraphics[width=2\columnwidth]{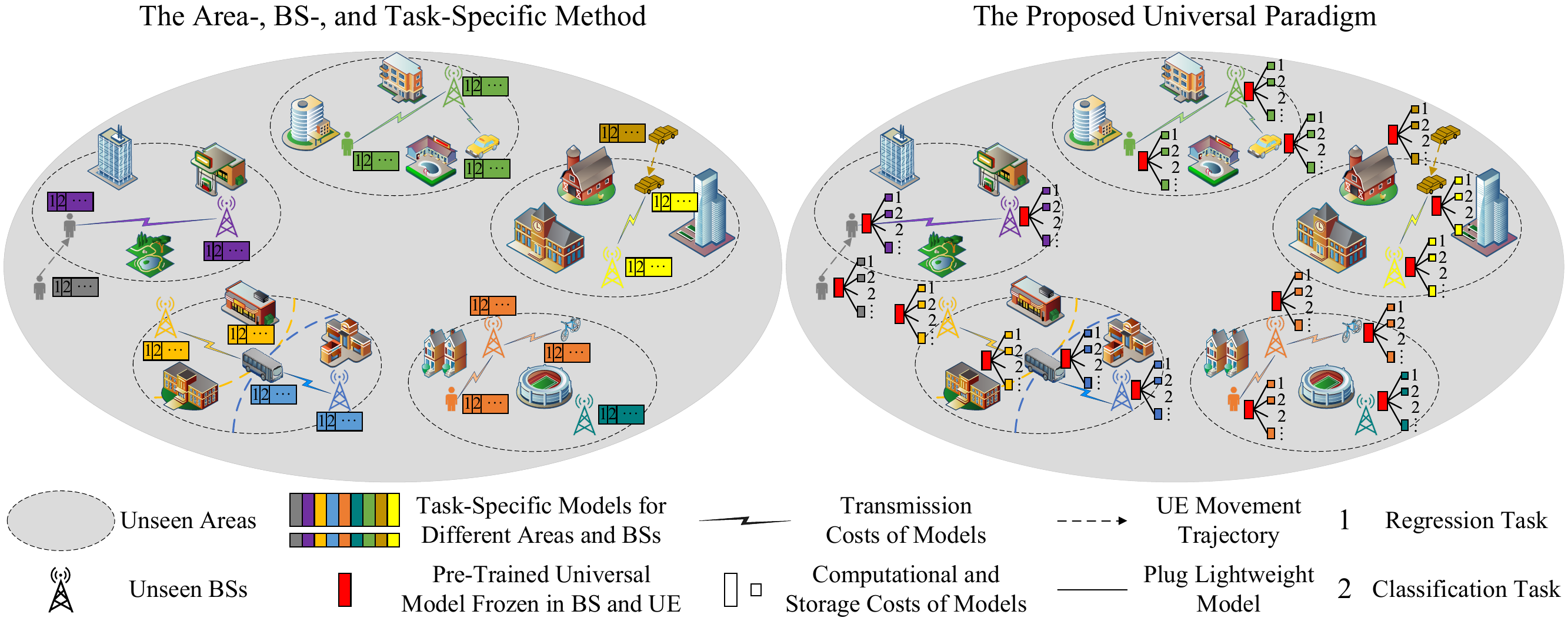}
\caption{The scenario- and task-specific method vs. the proposed universal paradigm.}
\label{f1}
\end{figure*}

\subsection{System Model}
This paper explores multiple dispersed areas of arbitrary size with varying building distributions, as illustrated in Fig. \ref{f1}. BSs are deployed at arbitrary positions within these areas. UEs enter new areas and access new BSs during their movement. Both BSs and UEs are involved in performing multiple tasks. Due to limited resources on UEs, BSs act as training entities and deliver ML models to the accessed UEs. The overall system costs include the computational, storage, and transmission resources allocated to all ML models across all areas.

The traditional system design is depicted on the left in Fig. \ref{f1}. Due to the diversity of building distributions, BS positions, and task objectives, the conventional method requires training separate ML models for each area, BS, and task. However, these specific models lack generalization capability and cannot leverage correlated knowledge from other scenarios. As the number of areas, BSs, and tasks increases, the demand for data collection and the number of specific models grow exponentially, leading to significantly increased system costs and model management challenges.

The proposed novel paradigm is illustrated on the right in Fig. \ref{f1}. The red multi-modal universal model is pre-trained and frozen in both BSs and UEs. This model can effectively process multi-modal data from unseen scenarios and extract task-agnostic universal representations. These representations can be used for direct classification and sensing or transformed into specific task objectives by plugging in lightweight scenario- and task-specific models. With the pre-trained model, UEs only need to receive and update low-cost specific models for task completion, significantly enhancing system flexibility.

\subsection{Channel Model}
We consider a massive MIMO system operating in orthogonal frequency division multiplexing (OFDM) mode with $N_c$ subcarriers. The BS is equipped with $N_t$ antennas arranged in a uniform linear array (ULA), while the UE has a single antenna. Consequently, the wireless channel between the BS and the UE can be written as,
\begin{equation}
\label{eq1}
\mathrm{h}(f)=\sum_{i=1}^{N_{\mathrm{path}}}\alpha_{i} e^{-j2\pi f\tau_i} \mathrm{a}(\theta_i),
\end{equation}
where $f$ is the carrier frequency, $N_{\mathrm{path}}$ is the number of propagation paths, $\alpha_i$ is the amplitude attenuation, $\tau_i$ denotes the time delay, and $\theta_i$ represents the AoA of the $i$-th path. Moreover, $\mathrm{a}(\theta_i)$ is the array vector expressed as,
\begin{equation}
\label{eq2}
\mathrm{a}(\theta_i)=[1,e^{-j\beta\cos{\theta_i}},\cdots,e^{-j\beta(N_t-1)\cos{\theta_i}}]^T,
\end{equation}
where $\beta=2\pi df/c$, $d$ is the antenna spacing, and $c$ is the speed of light. Consequently, the CSI matrix $\mathrm{H} \in \mathbb{C}^{N_t \times N_c}$ can be defined as,
\begin{equation}
\label{eq3}
\mathrm{H}=[\mathrm{h}(f_1),\mathrm{h}(f_2),\cdots,\mathrm{h}(f_{N_c})],
\end{equation}
where $\{f_i \mid i=1,2,\cdots,N_c\}$ is the set of subcarrier frequencies.

\subsection{Equivalence Analysis of Propagation Path Information in Physical Environment and Wireless Channel Modalities}
To address the curse of scenario and task generalization by leveraging multi-modal integration capabilities, it is important to first identify common communication attributes like the natural semantic relationships between image and text. Although environments and task objectives change, the principles of electromagnetic propagation remain invariant \cite{r2}. Both data from the physical environment and the wireless channel contain key information that explicitly or implicitly defines propagation paths. Therefore, they can be considered as different modalities of paths. We denote the path parameters $\{\alpha_i,\tau_i,\theta_i \mid i=1, 2, \cdots, N_{\mathrm{path}}\}$ as $\Theta_\mathrm{path}$.

For the physical environment modality, based on the information such as BS and UE positions as well as area building distributions, all potential action points affecting signal propagation can be identified using deterministic ray-tracing techniques that account for direct propagation, reflection, and diffraction models \cite{r32} \cite{r40}. Consequently, all possible propagation paths can be determined. The total physical distance of each path correlates with attenuation $\alpha_i$ and delay $\tau_i$, while the direction of the last hop for these paths correlates with AoA $\theta_i$. Thus, the path information $\Theta_\mathrm{path}$ can be extracted from the physical environment data, denoted as $\mathrm{D}$. The correlation can be expressed as follows,
\begin{equation}
\label{eq4}
\Theta_\mathrm{path} = f_{pe}(\mathrm{D}),
\end{equation}
where $f_{pe}(\cdot)$ represents the function that maps $\mathrm{D}$ to $\Theta_\mathrm{path}$. Therefore, $\mathrm{D}$ encapsulates both explicit environmental information and implicitly extractable path information.

For the wireless channel modality, according to Eqs. (\ref{eq1}), (\ref{eq2}), and (\ref{eq3}), aside from the configuration information of multi-antenna and multi-subcarrier systems, $\Theta_\mathrm{path}$ is inherently embedded within the channel data $\mathrm{H}$, such as CSI. Various algorithms, including multiple signal classification (MUSIC) \cite{r43}, estimation of signal parameters via rotational invariance techniques (ESPRIT) \cite{r44}, discrete fourier transform (DFT) \cite{r11}, and space-alternating generalized expectation-maximization (SAGE) \cite{r45}, can be employed to infer specific path parameters. The correlation can be expressed as follows,
\begin{equation}
\label{eq5}
\Theta_\mathrm{path} = f_{wc}(\mathrm{H}),
\end{equation}
where $f_{wc}(\cdot)$ represents the mapping function, which leverages these algorithms to extract $\Theta_\mathrm{path}$ from $\mathrm{H}$. Hence, $\mathrm{H}$ contains explicit path information.

In summary, the mapping functions $f_{pe}(\cdot)$ and $f_{wc}(\cdot)$ in Eqs. (\ref{eq4}) and (\ref{eq5}) not only exist but also indicate that propagation path information serves as the common semantic foundation that makes cross-modal comprehension feasible for bridging multiple communication modalities, such as the physical environment and the wireless channel. In real-world scenarios, signal propagation is also influenced by the scattering environment, and a one-to-one correspondence between the environment and channel modality samples remains valid.

\section{The Proposed Multi-Modal Pre-Training and Downstream Task Adaptation Paragidm}
This section introduces the details of the proposed multi-modal paradigm, which consists of two stages. The first stage involves pre-training for multi-modal alignment, achieving cross-scenario capability and extracting task-agnostic universal representations. Based on the original two-tower model, EPNN and CFENN, two specially designed components, CRNN and MSNN, are added. The second stage involves either direct ZSL or pluggable TOFT, which aims to flexibly and cost-effectively perform multiple downstream tasks across various unseen scenarios using the frozen pre-trained universal model.

\subsection{Pre-Training for Environment-Channel Modality Alignment}

\begin{figure*}[tb]
\centering
 \includegraphics[width=2\columnwidth]{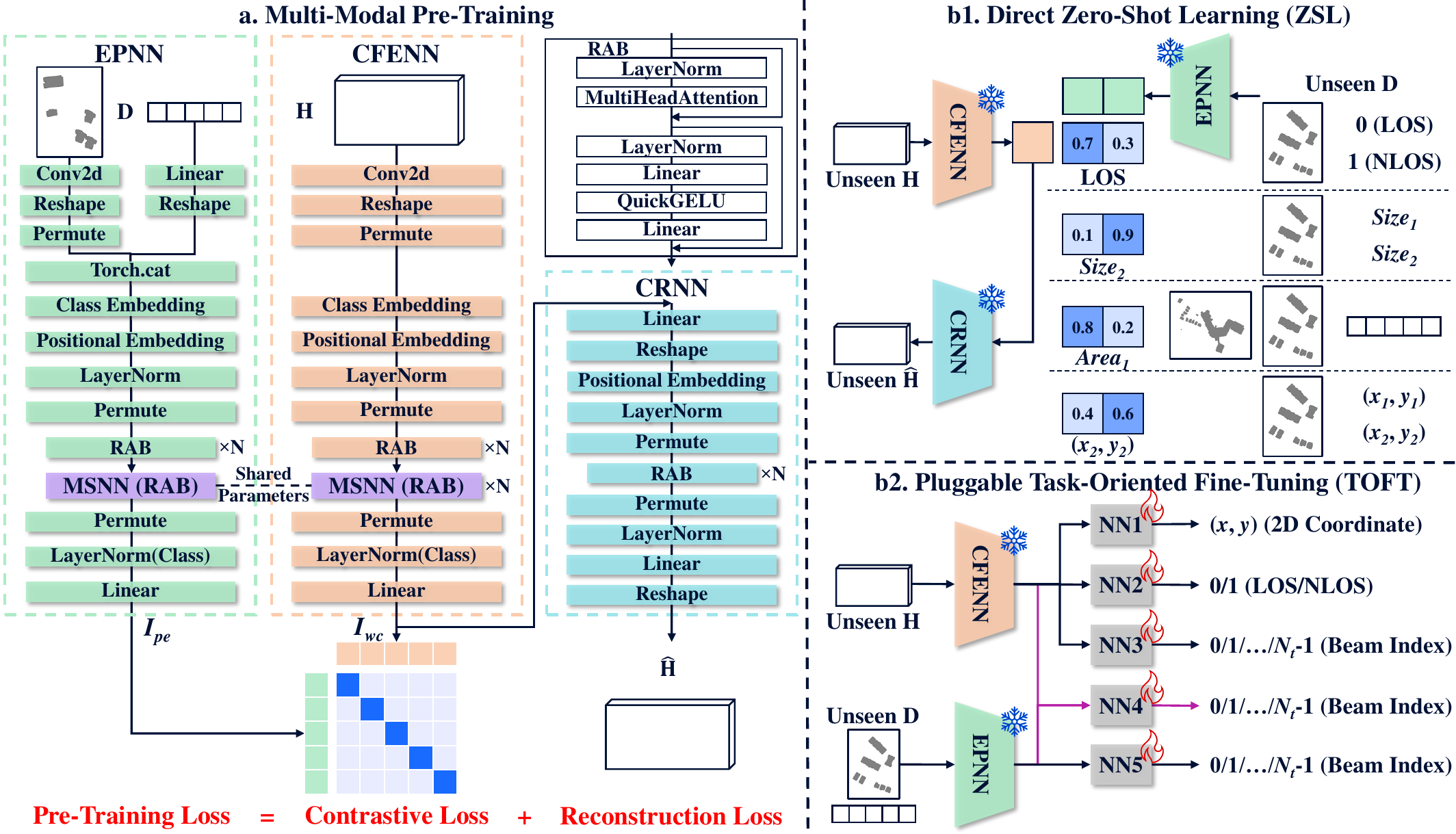}
\caption{Overview of the proposed 6G-oriented two-stage multi-modal pre-training and downstream task adaptation paradigm.}
\label{f3}
\end{figure*}

\textit{1) Two-Tower Model:} As mentioned above, due to the equivalence of path information in different communication modalities, we propose a two-tower model consisting of the green tower, EPNN, and the orange tower, CFENN, to extract features from the physical environment data $\mathrm{D}$ and the wireless channel data $\mathrm{H}$, respectively, as shown on the left in Fig. \ref{f3}.

The green EPNN with the transformer-based architecture \cite{r25} transforms the environment data $\mathrm{D}$ into latent representations $I_{pe}$ with dimensions of $\mathrm{Batch\_Size} \times \mathrm{Embedding\_Dim}$. Since $\mathrm{D}$ includes, but is not limited to, matrices like geographic maps and structured vectors such as BS and UE positions, feature extraction modules like the conv2d layer and the linear layer are separately employed to transform them into multiple patches, each being a high-dimensional vector. If other data formats, such as 3D-structured spatial maps are integrated, modules like the conv3d layer can be utilized instead. These two patch sequences are then concatenated for subsequent feature fusion. An extra learnable identifier, i.e., the class patch, is added before the fused patch sequence by the class embedding layer for global feature information aggregation. Subsequently, positional encoding is applied to each patch to provide patch position information via the positional embedding layer. The permute layers reorder the $\mathrm{Batch\_Size}$, $\mathrm{Patch}$, and $\mathrm{Embedding\_Dim}$ dimensions of the data to meet the input format requirements of subsequent layers. The patch sequence is then fed into residualattentionblocks (RABs), where attention weights among patches are calculated to capture dependencies by the multiheadattention layer, and non-linearity is introduced by the linear and quickGELU layers. Finally, the class patch is transformed into the embedding space.

The orange CFENN with the vision transformer (ViT)-based architecture \cite{r21} transforms the channel data $\mathrm{H}$ into latent representations $I_{wc}$ with dimensions of $\mathrm{Batch\_Size} \times \mathrm{Embedding\_Dim}$. The CSI matrix is sequentially segmented into small antenna-subcarrier patches by the conv2d layer. A class patch is added before the patch sequence, and positional encoding is applied to each patch. The patch sequence is then fed into RABs. Finally, the class patch is transformed into the embedding space.

The feature dimensions of $I_{pe}$ and $I_{wc}$ are set to be the same, aiming to map both modality representations to a same embedding space. Multi-modal alignment can then be conducted using contrastive learning \cite{r1}, which intends to acquire representations with discriminative features by drawing related environment-channel modality pairs close together and pushing unrelated pairs far apart in the embedding space. Firstly, the $\mathrm{Embedding\_Dim}$-dimensional modality representation of each sample in $I_{pe}$ and $I_{wc}$ is normalized individually. Then, the cosine similarity matrices between samples of $I_{pe}$ and $I_{wc}$, as well as between $I_{wc}$ and $I_{pe}$, can be calculated by
\begin{equation}
M_{pe\_wc}=I_{pe}{I_{wc}}^{T},
\end{equation}
\begin{equation}
M_{wc\_pe}=I_{wc}{I_{pe}}^{T},
\end{equation}
where $T$ represents the matrix transpose. The ground truth $G$ for contrastive learning is an identity matrix, as shown in the lower left of Fig. \ref{f3}. The dimensions of $M_{pe\_wc}$, $M_{wc\_pe}$, and $G$ are $\mathrm{Batch\_Size} \times \mathrm{Batch\_Size}$. By this method, the similarities of related environment-channel pairs tend to be maximized, while the similarities of unrelated pairs tend to be zero to ensure clear differentiation. The contrastive loss functions for the similarity matrices and ground truth are calculated as
\begin{equation}
L_{pe\_wc}=\mathrm{Contrastive\ Loss} (M_{pe\_wc},G),
\end{equation}
\begin{equation}
L_{wc\_pe}=\mathrm{Contrastive\ Loss} (M_{wc\_pe},G).
\end{equation}
The final contrastive loss is the average of the above two terms and is written as
\begin{equation}
L_C=\frac{L_{pe\_wc}+L_{wc\_pe}}{2}.
\end{equation}
With the introduced two-tower model architecture and contrastive loss function, the modality gap between the physical environment and the wireless channel can be reduced. Since multi-modal data includes, but is not limited to, area, BS, and task information, generalization across scenarios and tasks can be anticipated. The extracted task-agnostic multi-modal representations encapsulate rich high-level knowledge, contributing to the AI-native evolution of communication systems.

\begin{figure*}[t]
    \hspace{-0.4cm}
    \subfigure[t-SNE of Two-Tower Model]{
		\begin{minipage}[b]{0.3\textwidth}
	    \centering
		\includegraphics[width=0.89\textwidth]{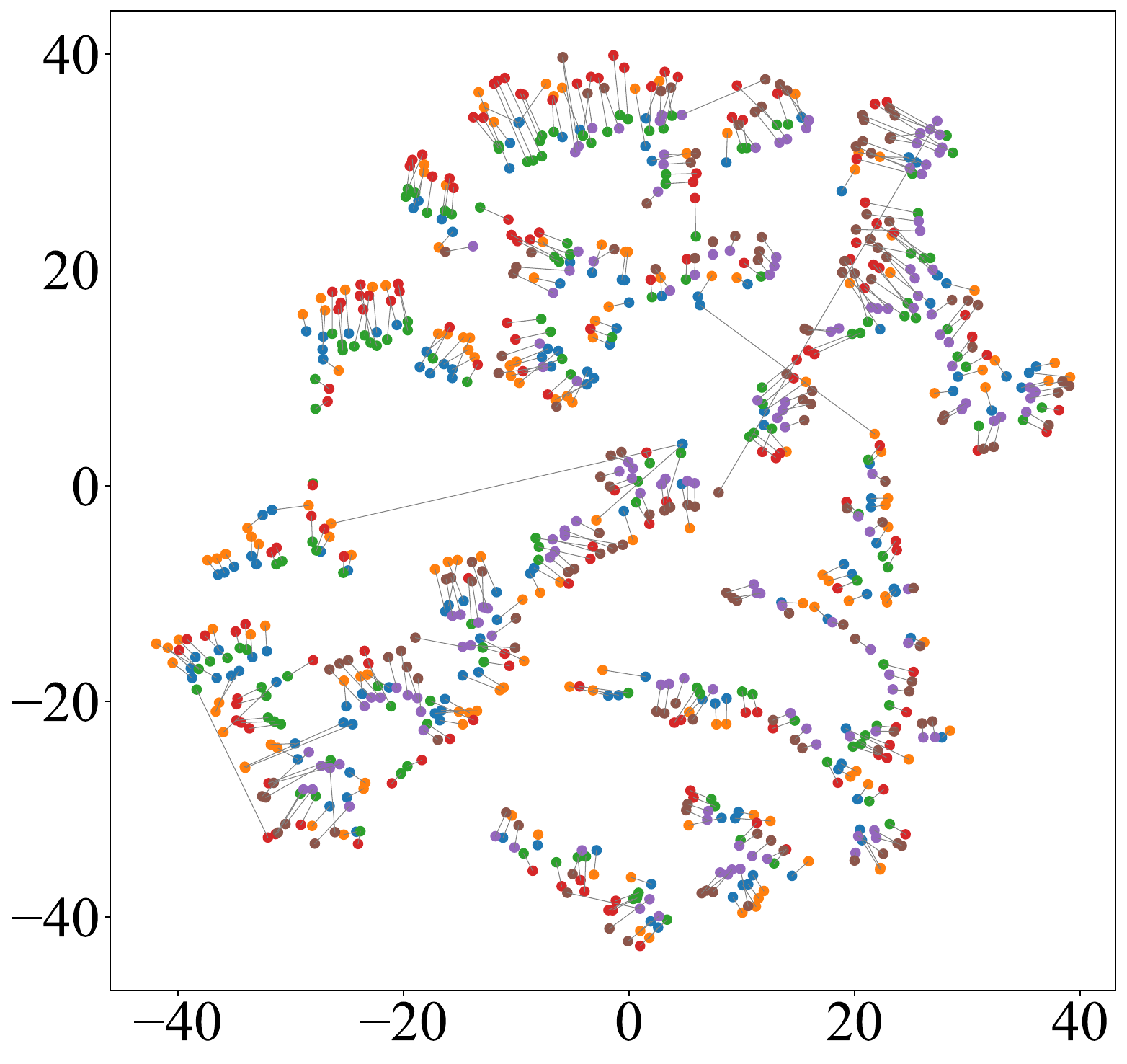} 
        \end{minipage}
    }
    \hspace{-1.2cm}
	\subfigure[UMAP of Two-Tower Model]{
		\begin{minipage}[b]{0.3\textwidth}
		\centering
   	 	\includegraphics[width=0.864\textwidth]{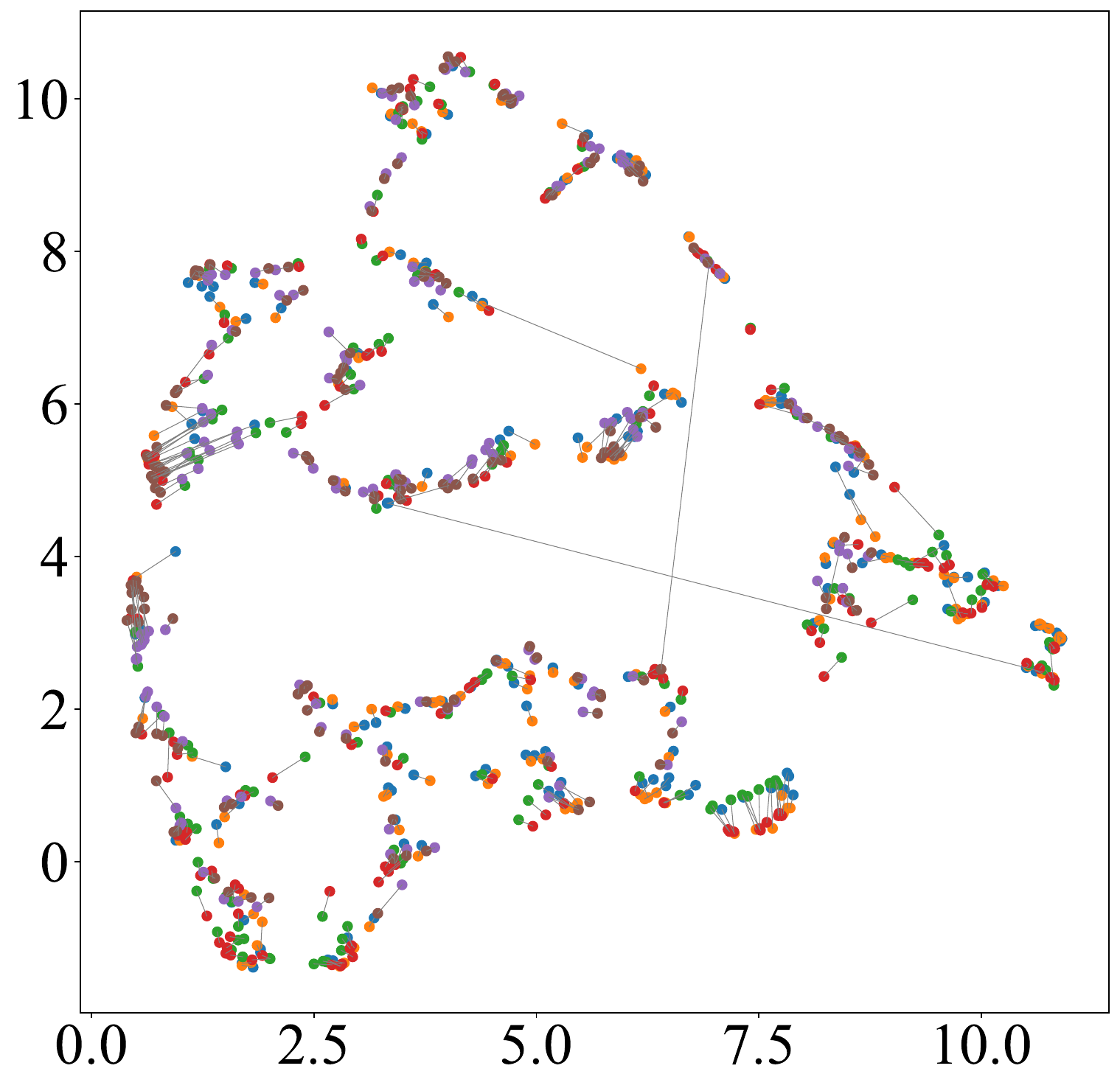}
		\end{minipage}
	}
    \hspace{-0.95cm}
	\subfigure[PCA of Two-Tower Model]{
		\begin{minipage}[b]{0.3\textwidth}
		\centering
   	 	\includegraphics[width=1.586\textwidth]{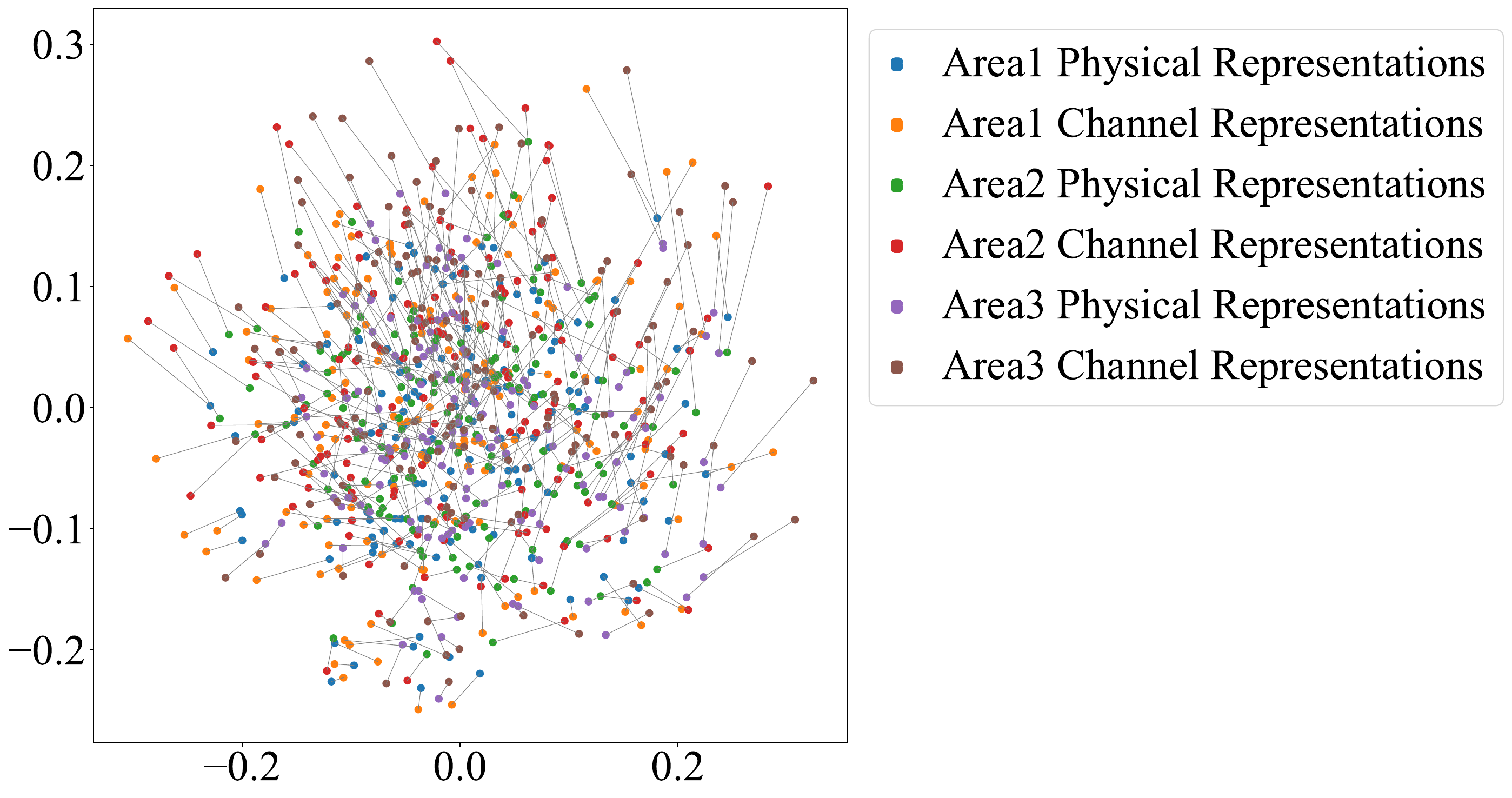}
		\end{minipage}
	}
	\\ 
	\subfigure[t-SNE after Adding CRNN]{
		\begin{minipage}[b]{0.3\textwidth}
			\includegraphics[width=0.89\textwidth]{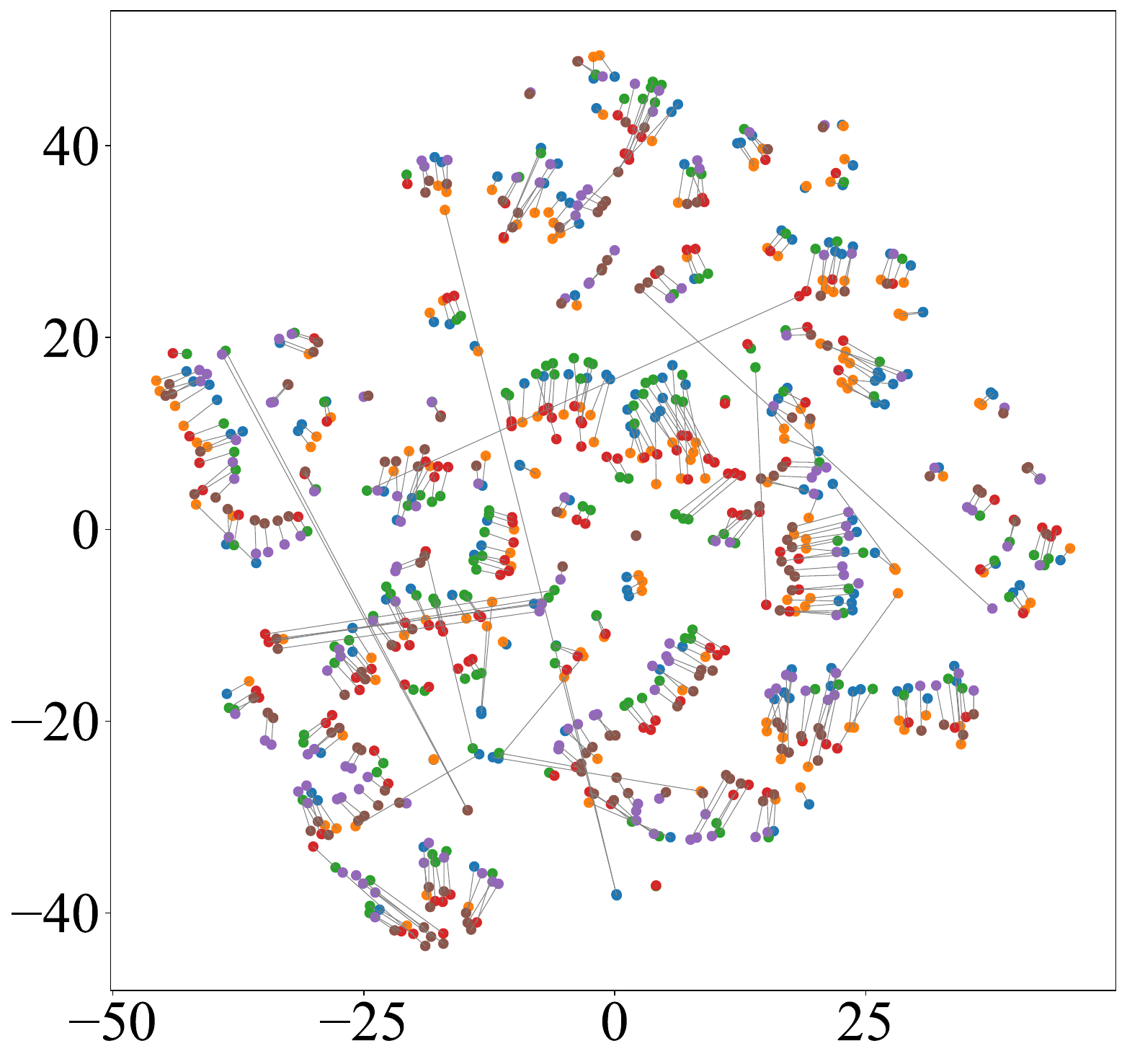} 
		\end{minipage}
	}
    \hspace{-1.1cm}
	\subfigure[UMAP after Adding CRNN]{
		\begin{minipage}[b]{0.3\textwidth}
	 	\includegraphics[width=0.864\textwidth]{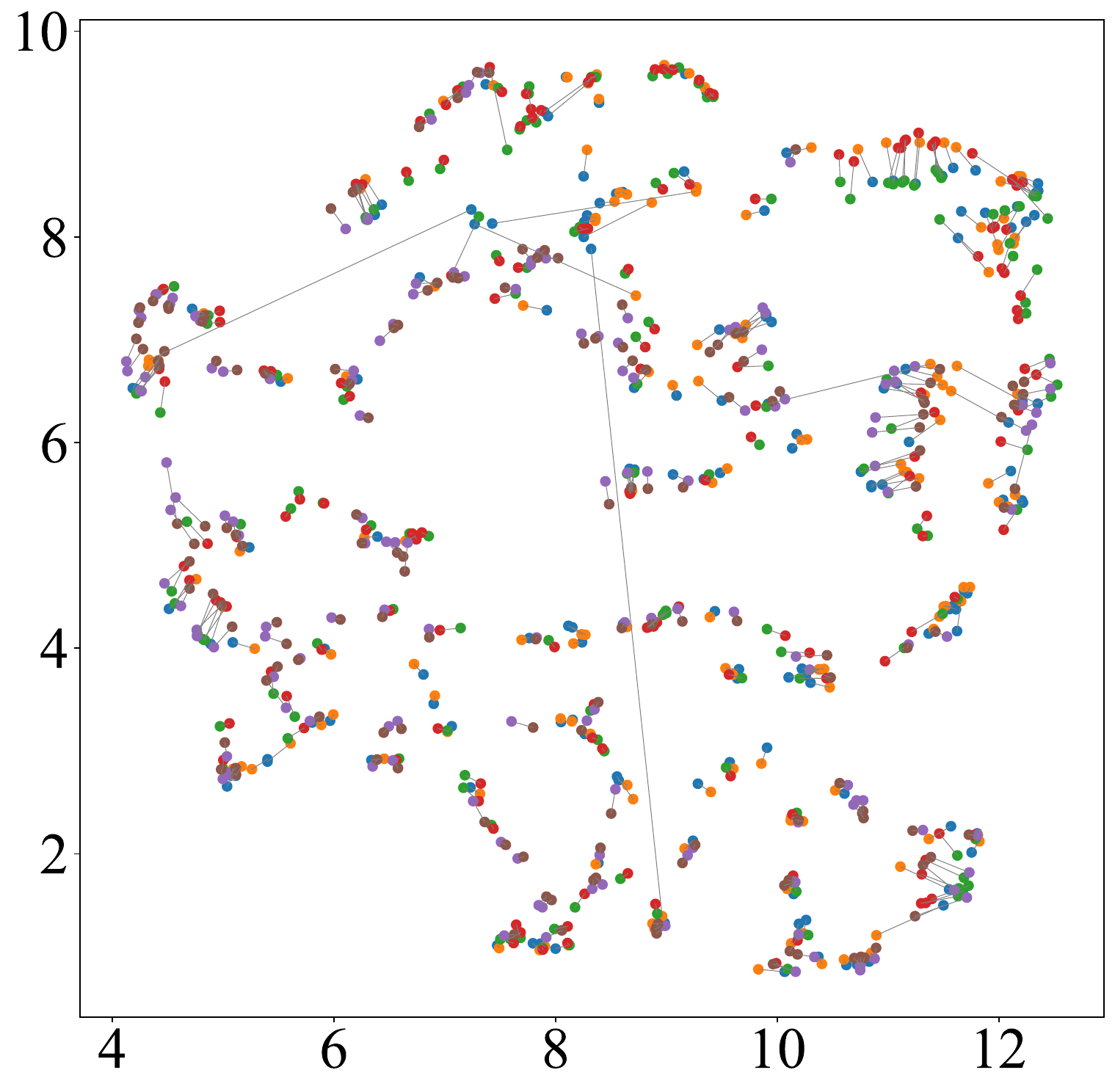}
		\end{minipage}
	}
    \hspace{-1.35cm}
	\subfigure[PCA after Adding CRNN]{
		\begin{minipage}[b]{0.3\textwidth}
	 	\includegraphics[width=1.586\textwidth]{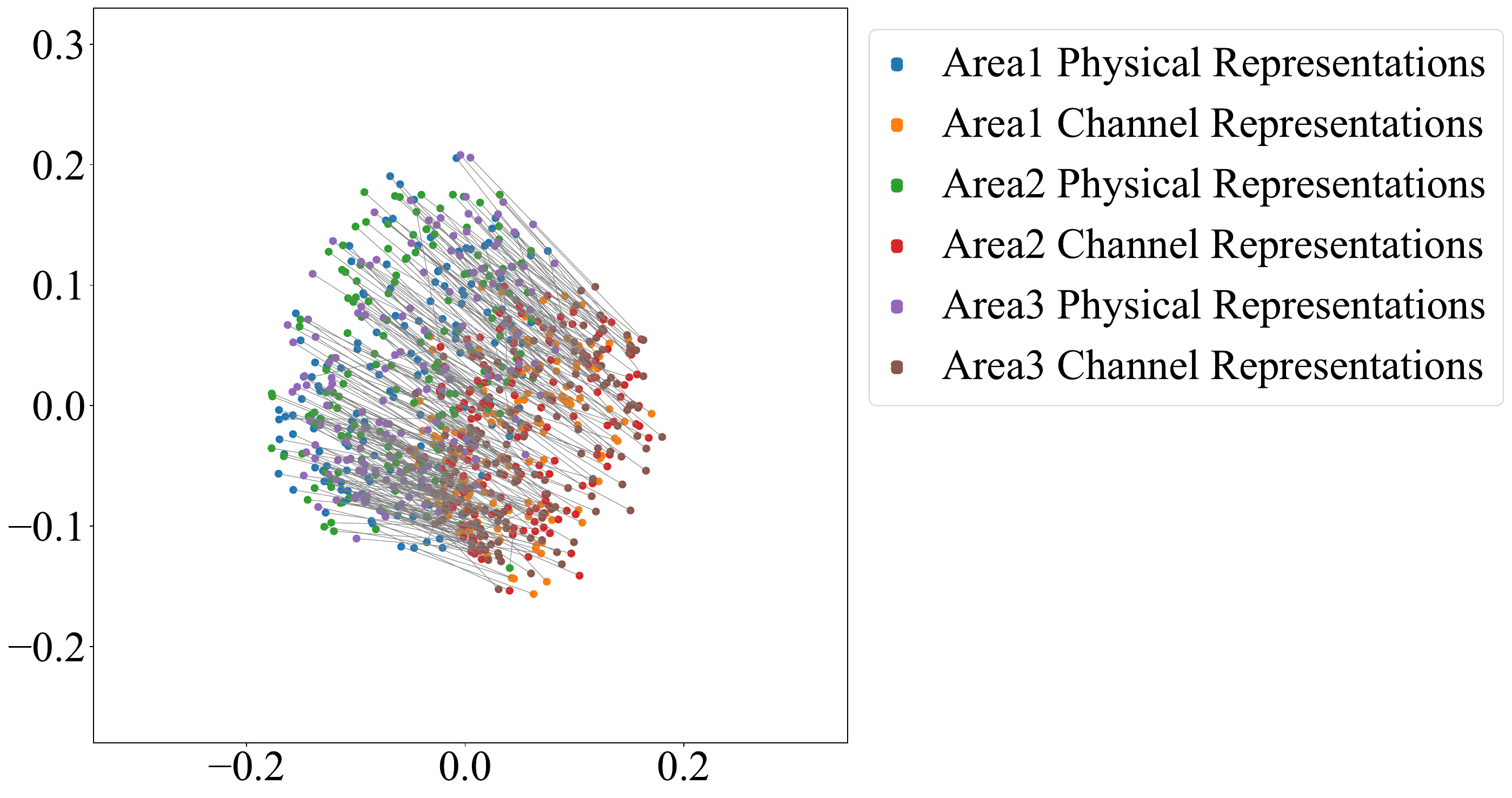}
		\end{minipage}
	}
	\caption{Visualization of the embedding space of the pre-trained two-tower model and after adding CRNN using t-SNE, UMAP, and PCA methods.}
	\label{f4}
\end{figure*}

\textit{2) CRNN:} Given that contrastive learning can bridge multiple modalities, we want to further clarify the feature extraction and alignment capabilities of the two-tower model. Since only dimension-reduction EPNN and CFENN and a self-supervised learning method are utilized, without explicit task-specific labels for supervised training, it is crucial to ensure that the two-tower model adequately captures the features of multi-modal data. CSI feedback is a reconstruction-oriented compression task, which requires the latent channel representations to encompass all path information \cite{r34}. Therefore, we design a blue CRNN to evaluate the reconstructability of these low-dimensional representations, as depicted on the left in Fig. \ref{f3}.

First, the wireless channel representations are upscaled by a linear layer, and the obtained high-dimensional vector is reshaped into a patch sequence. Positional encoding is added to each patch, and the patch sequence is fed into RABs. Finally, the sequence is transformed into the reconstructed CSI, denoted as $\hat{\mathrm{H}}$. The CRNN is trained in a supervised manner to recover the original CSI from the channel representations of the two-tower model. The reconstruction loss is measured by the mean squared error (MSE), calculated as follows,
\begin{equation}
L_R=\frac{1}{N_{\mathrm{H}}}\sum_{i=1}^{N_{\mathrm{H}}}\|\hat{\mathrm{H}}_i-\mathrm{H}_i\|_2^2,
\end{equation}
where $\|\cdot\|_2$ denotes the Euclidean norm, and $N_{\mathrm{H}}$ is the number of CSI samples. We observed that the reconstruction loss did not converge. However, when joint training the same CFENN and CRNN from scratch, the reconstruction loss did converge. This indicates that the channel representations extracted by the two-tower model drift or lose propagation path information.

Furthermore, we visualize the embedding space of the pre-trained two-tower model to observe the interrelationships of both modality representations using three manifold learning methods \cite{r26}: t-distributed stochastic neighbor embedding (t-SNE), uniform manifold approximation and projection (UMAP), and principal component analysis (PCA), as shown in Fig. \ref{f4}-(a), (b), and (c). The t-SNE and UMAP results show that most modality pairs are clearly distinguished, with only a few pairs connected by long lines not being well-aligned. In contrast, the PCA result appears highly disordered. Since t-SNE and UMAP excel at capturing nonlinear relationships, whereas PCA is suited for exhibiting linear relationships, these results imply that the two-tower model's embedding space lacks linear relationships. Furthermore, this confirms that the two-tower model structure and contrastive loss function alone are insufficient to preserve the relevant information, which is subsequently needed for reconstructing $\mathrm{H}$ using CRNN.

It has been demonstrated that different initial weights would lead to inconsistent convergence directions in image-text two-tower models \cite{r22}, which is unreliable and risky for wireless systems. Therefore, our pre-training model introduces unique convergence guidance by adding the CRNN branch, which offers many advantages. For data collection, no extra overhead is introduced due to the self-supervised manner. For the CSI feedback task, it can be performed directly. For the justification of the embedding space, the wireless channel representations are compelled to encompass all propagation path information, thereby providing more comprehensive features to the associated physical environment modality. The final pre-training loss function comprises the multi-modal contrastive loss and the channel reconstruction loss, expressed as follows,
\begin{equation}
\label{eq12}
L=L_C+L_R.
\end{equation}
The visualization of the embedding space after incorporating the CRNN module is depicted in Fig. \ref{f4}-(d), (e), and (f). The t-SNE and UMAP results resemble those of the two-tower model, indicating that the nonlinear relationships are unaffected. However, the PCA result shows that the directions of the lines connecting both modalities are consistent, marking the emergence of linear relationships in the embedding space.

\textit{3) MSNN:} For multi-modal models, modality interaction and fusion can be achieved not only through loss functions, such as the introduced contrastive loss, but also by the model architecture \cite{r23}\cite{r24}. Due to the structured forms and distinct meanings of communication data from different modalities, modality-specific NNs, such as the proposed EPNN and CFENN, are designed and trained for single modalities and cannot be used for other modalities. Developing a unified NN module MSNN with modality-agnostic parameters and weights to separately extract features from multiple modalities offers significant advantages. Firstly, the MSNN module is more parameter-efficient, as the same set of parameters can be utilized for various modalities. Secondly, the MSNN is trained using data from all modalities, facilitating the integration and complementation of knowledge from different modalities. This helps achieve robust performance across various downstream tasks.

Building the MSNN places stringent requirements on model structure design. Unlike other classical architectures limited to specific data formats and scenarios, such as convolutional neural network (CNN) and recurrent neural network (RNN), transformer and its attention mechanism \cite{r25} have demonstrated versatility across various modalities, such as language, vision, and audio, showing potential for achieving unified multi-modal intelligence. Recently, unified frameworks in the CS community, such as BEiT-3 \cite{r27} and Meta-Transformer \cite{r28}, have utilized transformer-based modality-shared NNs to enhance the understanding capabilities of multi-modal models.

Since the wireless channel representations extracted by the CFENN contain all path information after adding the CRNN, the EPNN is also directed to extract environment and path information from the physical environment modality data through the indirect intra-modal and direct inter-modal contrastive loss. Therefore, we design a purple MSNN (RABs) to further fuse information from different modalities, as depicted on the left in Fig. \ref{f3}. The modality-specific parts in the EPNN and CFENN are used to transform their respective modality data into intermediate features that the MSNN can process. The pre-training loss function remains as in Eq. (\ref{eq12}). The proposed MSNN stands out as an innovative design for the efficient processing of multi-modal communication data.

\subsection{Downstream Task Adaptation}
Leveraging the generalization capability and learned semantic knowledge of the frozen pre-trained multi-modal universal model, two flexible and cost-effective downstream task adaptation methods, direct ZSL and pluggable TOFT, are employed to perform various communication and sensing, classification and regression sub-tasks in unseen scenarios with minimal or no tuning parameters, as illustrated on the right in Fig. \ref{f3}.

\textit{1) ZSL:} Leveraging the alignment of various modalities achieved through pre-training, the ZSL method can directly select the correct option from multiple unseen and uncertain choices of one modality by comparing the similarities between these option features with knowledge from other modalities, thereby enabling effective classification and sensing functions, as depicted in Fig. \ref{f3}-b1. The advantages of ZSL are that it requires no additional refinement and can select from an arbitrary number of samples, making it a promising solution for the low-latency and high-dynamics demands of future 6G systems. For instance, the UE's line-of-sight (LOS) status can be immediately identified based on estimated CSI. The unseen geographic map and binary indicators (0 for LOS and 1 for non-line-of-sight (NLOS)), as well as unseen CSI, are fed into the frozen EPNN and CFENN, respectively. Based on the output $I_{pe1}$, $I_{pe2}$, and $I_{wc1}$, the similarities between representations of CSI and LOS option, as well as CSI and NLOS option, can be calculated respectively. The class with a higher probability is the prediction result, which can be expressed as,
\begin{equation}
R_\mathrm{los}=\mathrm{argmax}(I_{pe1}{I_{wc1}}^T, I_{pe2}{I_{wc1}}^T).
\end{equation}
If additional information, such as BS and UE positions, is available and fed into the pre-trained model, the classification accuracy would be further improved.

Another example is environment sensing. When area characteristics, such as physical sizes or building distributions, are uncertain, they can be inferred from several options using the CSI and position information of the UE in the area. After pre-training, transceivers can generalize from BS and UE to other devices, such as vehicles. This capability enables the system to identify the actual communication device from multiple possible positions, akin to a multi-user matching task \cite{r10}. Furthermore, the frozen CFENN and CRNN can directly compress and recover unseen CSIs in new scenarios.

\textit{2) TOFT:} The TOFT method involves plugging lightweight area-, BS-, and task-specific NNs after the frozen pre-trained model for supervised training, transforming universal modality representations into specific sub-task objectives, such as UE position, LOS status, and beam index, as depicted in Fig. \ref{f3}-b2. The advantage of TOFT is that only minimal tuning parameters are required, resulting in faster model convergence and lower system costs. More importantly, because the multi-modal universal model possesses original path propagation knowledge, TOFT can perform tasks whose objectives were not pre-trained, such as beamforming codeword selection.

In practical applications, given the challenges of data collection in complex scattering environments, many tasks can be performed using different inputs, such as CSI/position-based beamforming \cite{r29}\cite{r30}\cite{r31}. However, due to the different information integrity of CSI and position data, the difficulty of extracting equivalent features from them varies, leading to distinct task performance. Since our paradigm aligns various modalities, the EPNN can provide more comprehensive path information for the position input, thereby achieving higher accuracy compared to traditional methods. Moreover, considering the consistent dimensions of different modality representations, a modality-shared fine-tuning network NN4 capable of processing all these features can be trained to alleviate the burden caused by multiple modality-specific NNs.

\section{Experimental Results and Analysis}
This section details the dataset construction, evaluates our multi-modal universal model, and compares the performance of ZSL and TOFT against multiple benchmarks across exemplary downstream tasks in previously unseen areas and BSs.

\subsection{Dataset Generation}
The wireless AI research dataset (WAIR-D) is employed in the experiments, where radio propagation paths are generated using a ray-tracing simulator with specified environmental configurations \cite{r32}. The dataset randomly chooses 10,000 real-world areas of varying sizes from over 40 cities and provides their building layout information. WAIR-D comprises scenario 1 (S1) and scenario 2 (S2). In S1, there are 10,000 areas, each containing 5 BSs and 30 sparsely distributed UEs. In S2, 100 areas are selected from the 10,000 areas, each containing 1 BS and 10,000 densely deployed UEs. For pre-training, samples from 9,000 areas numbered \#01001 to \#10000 in S1 and 90 areas from the same set in S2 are used, allowing the pre-training model to capture extensive area characteristics and improve precision within each area. For downstream task adaptation, unseen samples from the remaining 1,000 areas numbered \#00001 to \#01000 in S1 and 10 areas from the same set in S2 are utilized. Based on the area sample numbers in S1 and S2, samples from S1 are used for ZSL, while samples from S2 are used for both ZSL and TOFT. The parameters of the WAIR-D dataset are summarized in Table \ref{t1}.

\begin{table}[t]
\caption{Parameters of the WAIR-D Dataset}
\begin{center}
\begin{tabular}{ll}
\hline
\textbf{Parameter}&\textbf{Value}\\
\hline
Carrier Frequency&28GHz\\
Bandwidth&46.08MHz\\
Sub-Carrier Number&64\\
Antenna Configuration&32 ULA Tx Ports, 1 Rx Port\\
BS and UE Height&6m, 1.5m\\
Pre-Training Dataset & $9000\times5\times30+90\times1\times10000$ samples\\
ZSL and TOFT Dataset & $1000\times5\times30+10\times1\times10000$ samples\\
\hline
\end{tabular}
\label{t1}
\end{center}
\end{table}

The pre-training data generated by WAIR-D includes BS and UE positions, UE LOS status, physical-to-pixel scaling factor, area top-view map, and CSI. In practice, these data can be obtained from channel measurement campaigns \cite{r42} and public sources such as \textit{OpenStreetMap}, satellite images \cite{r41}, and navigation applications. Therefore, this paper does not need data that is difficult to acquire.

\begin{figure}[tb]
\centering
\includegraphics[width=1\columnwidth]{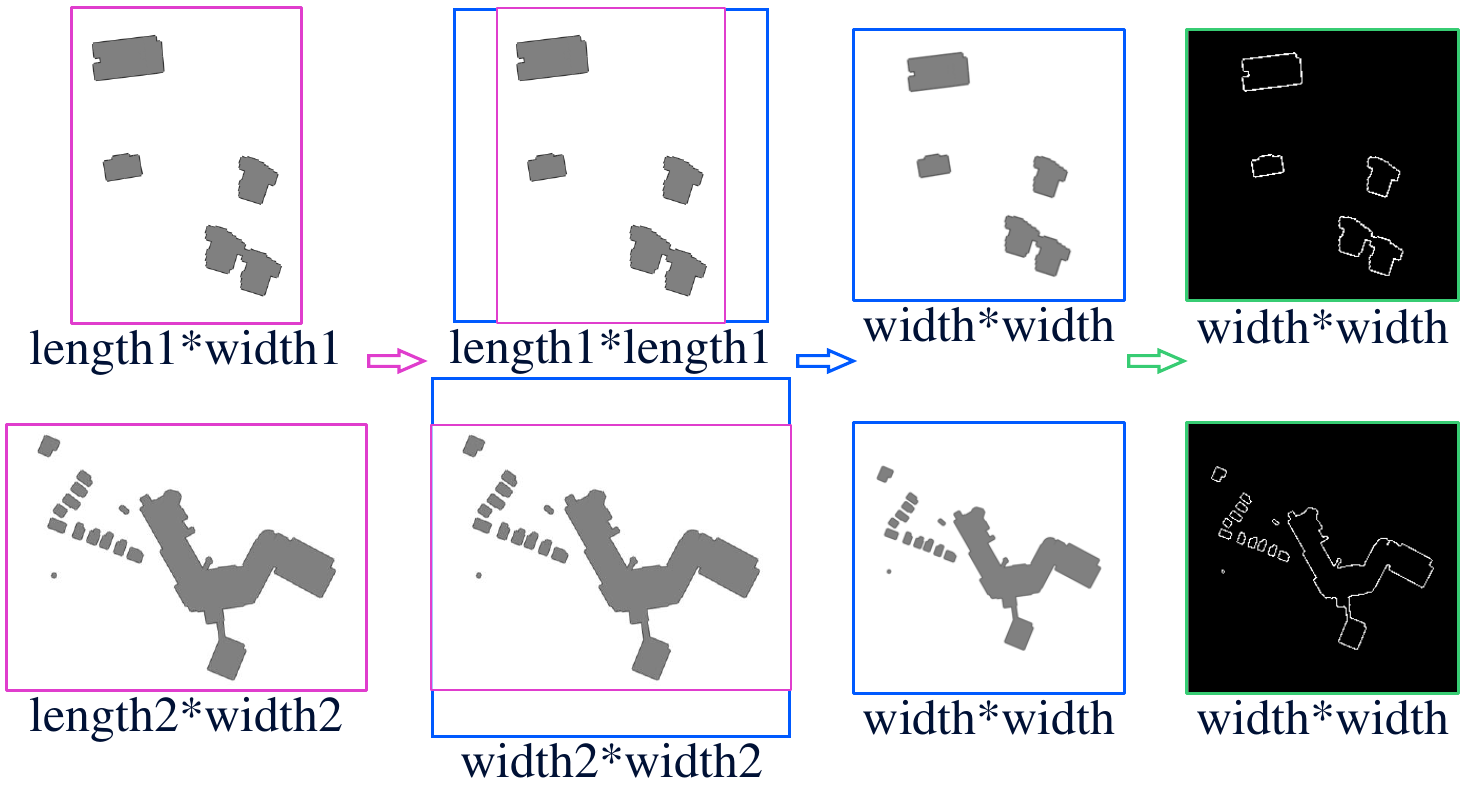}
\caption{From left to right: top-view maps of different-sized physical areas, filling maps into square shapes, scaling square top-view maps to uniform-size pixel images, transforming pixel images into binary building edge images.}
\label{f2}
\end{figure}

For all areas, buildings exceed both BSs and UEs in height. The entire area map rather than BS/UE-centered local map is adopted. Since ML models commonly handle data of uniform sizes, physical area top-view maps of varying sizes are transformed into square pixel images of the same size without altering their spatial characteristics, as depicted in Fig. \ref{f2}. The scaling factor for this transformation can be extracted accordingly. Due to the critical effect of building surfaces on signal propagation \cite{r33}, edge extraction (the last step in Fig. \ref{f2}) helps reduce data complexity and direct model attention.

The 3D BS and UE positions, LOS status, and scaling factor are organized into an 8-dimensional vector. The planar physical coordinates of BSs and UEs are translated and scaled into pixel coordinates using the same scale as their respective area maps to maintain consistency. The LOS status not only corroborates whether the buildings in area maps obstruct the link between the BS and UE positions but also enables the pre-trained universal model to quickly ascertain the LOS status of unseen UEs in new scenarios. Scaling factor is the key to bridging pixel map and positions with CSI. The complex channel matrix is reshaped into a real matrix by concatenating its real and imaginary components. To facilitate model convergence, each reshaped CSI is normalized individually \cite{r33.5}.

\subsection{Pre-Training Model Settings and Evaluation}
The detailed structure of the pre-training model is depicted in Fig. \ref{f3}-a, with approximately $19.1$M parameters. The cross-entropy function is employed as the contrastive loss. The universal embedding space dimension is $128$. AdamW is utilized as the optimization algorithm, with a Cosine Annealing Warmup Restarts scheduler that adjusts the learning rate (LR) from 0 to 5e-4. The model is pre-trained for 100 epochs with a batch size of 64. The effectiveness of the pre-trained multi-modal universal model is evaluated using the following ZSL methods, which assess its precision in distinguishing unseen environment and channel data. This approach is commonly used in contrastive learning models \cite{r1}.

\begin{figure}[t]
    \subfigure{
		\begin{minipage}[b]{0.2\textwidth}
	    \centering
		\includegraphics[width=1\textwidth]{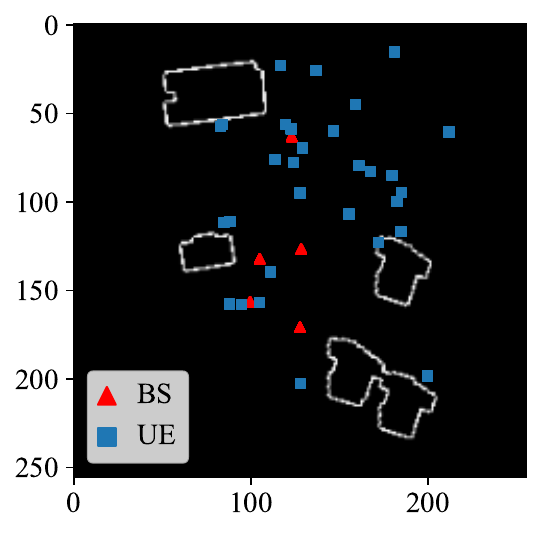} 
        \end{minipage}}
    \hspace{-0.3cm}
	\subfigure{
		\begin{minipage}[b]{0.29\textwidth}
		\centering
   	 	\includegraphics[width=1\textwidth]{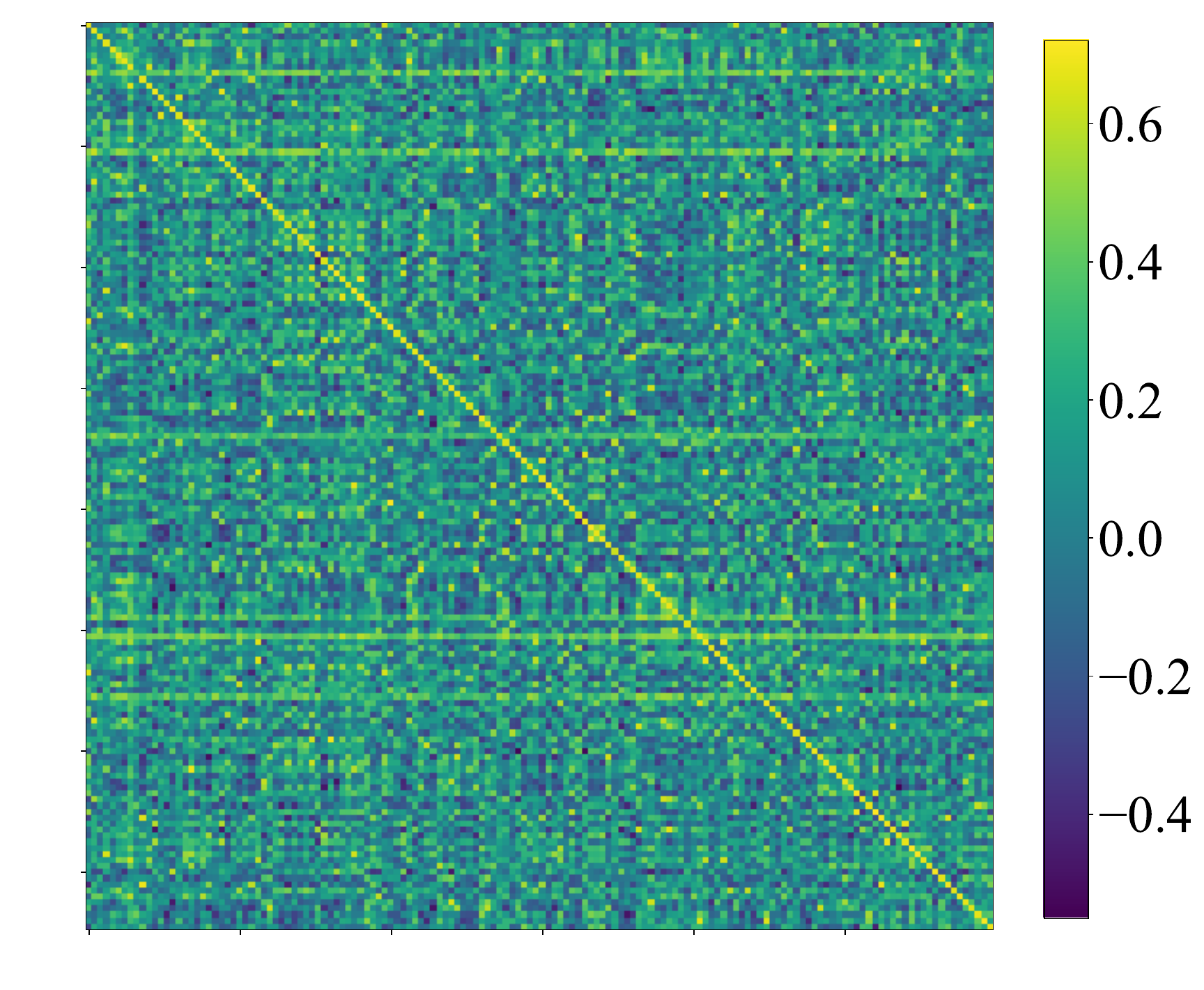}
		\end{minipage}}
	\caption{Left: 5 BSs and 30 UEs in area \#00032 in S1. Right: Similarity matrix of the corresponding 150 environment-channel modality representation pairs.}
	\label{f5}
\end{figure}

\begin{figure}[tb]
\centering
\includegraphics[width=1\columnwidth]{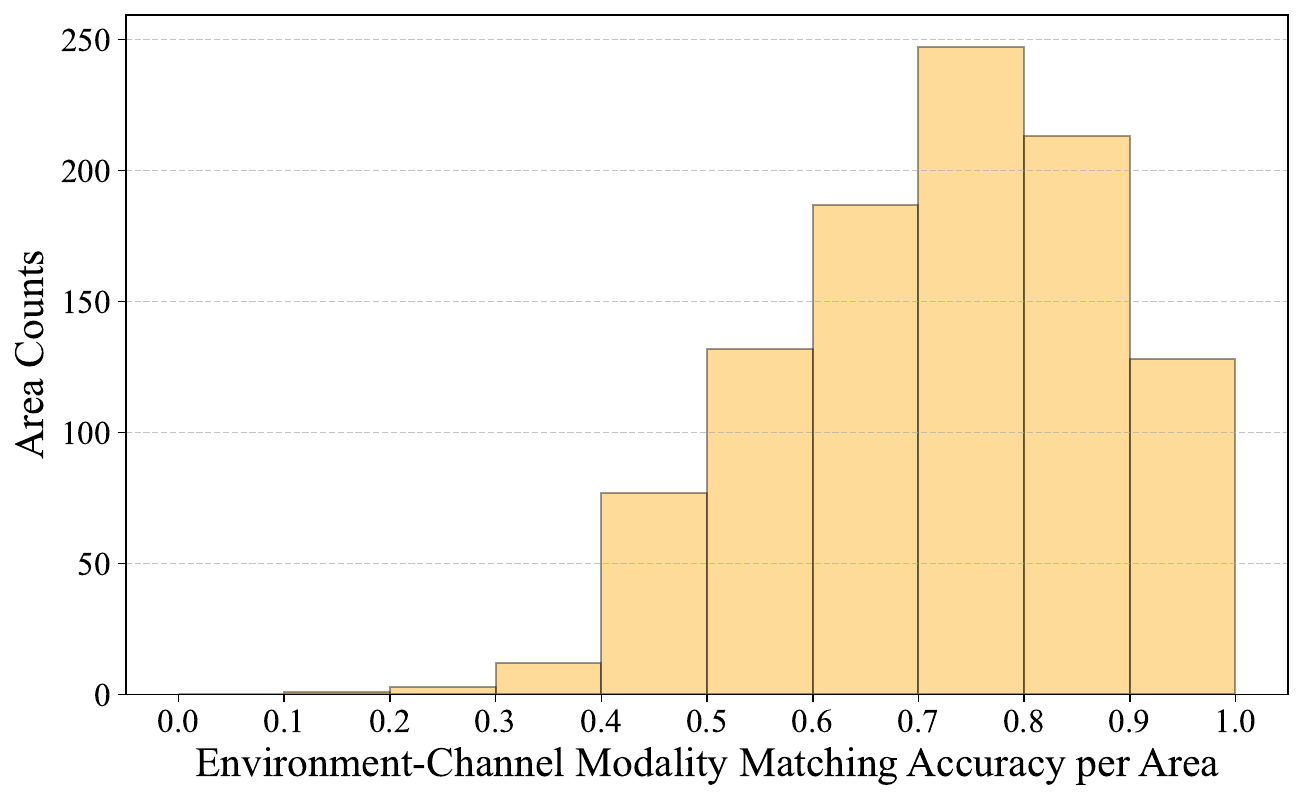}
\caption{Statistical analysis of multi-modal matching accuracy for the pre-trained universal model in 1,000 unseen S1 areas.}
\label{f5.5}
\end{figure}

\begin{figure}[tb]
    \subfigure{
		\begin{minipage}[b]{0.235\textwidth}
	    \centering
		\includegraphics[width=1\textwidth]{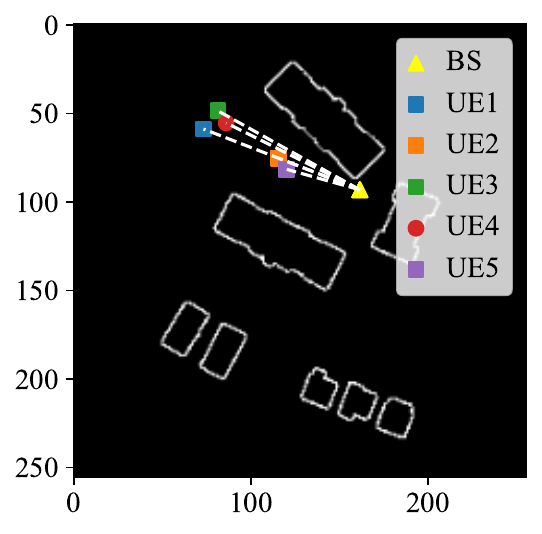} 
        \end{minipage}}
	\subfigure{
		\begin{minipage}[b]{0.235\textwidth}
		\centering
   	 	\includegraphics[width=1\textwidth]{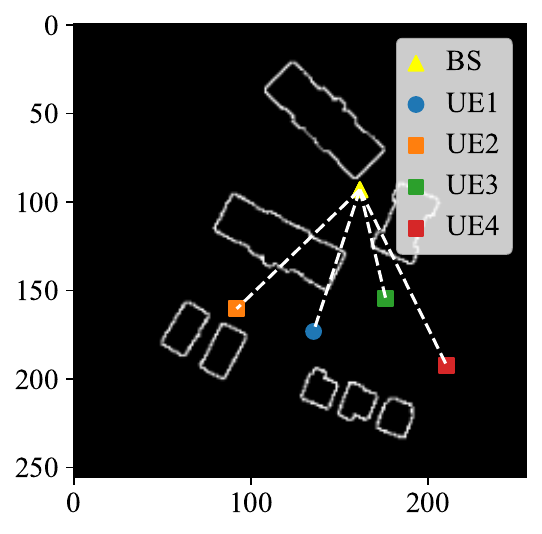}
		\end{minipage}}
	\caption{Left: We compare the similarities between the representations of 5 LOS UEs' positions and UE4's CSI in area \#00001, with probability distribution [1.348e-6, 2.023e-18, 1.677e-2, 0.983, 9.354e-17]. Right: We compare the similarities between the representations of 4 NLOS UEs' positions and UE1's CSI, with probability distribution [0.883, 0.058, 0.054, 0.005].}
	\label{f6}
\end{figure}

\begin{figure}[tb]
    \subfigure{
		\begin{minipage}[b]{0.235\textwidth}
	    \centering
		\includegraphics[width=1\textwidth]{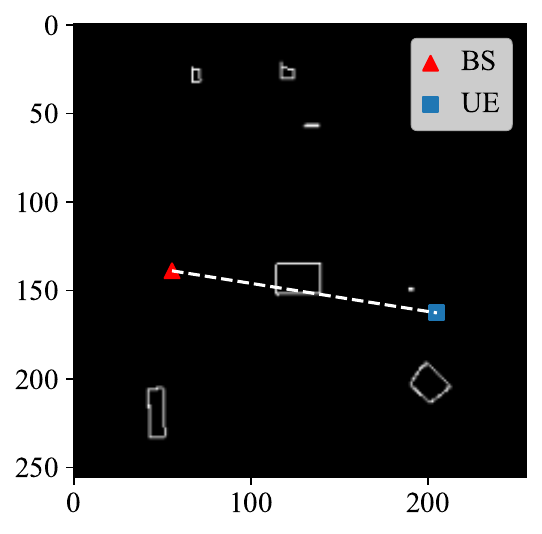} 
        \end{minipage}}
	\subfigure{
		\begin{minipage}[b]{0.235\textwidth}
		\centering
   	 	\includegraphics[width=1\textwidth]{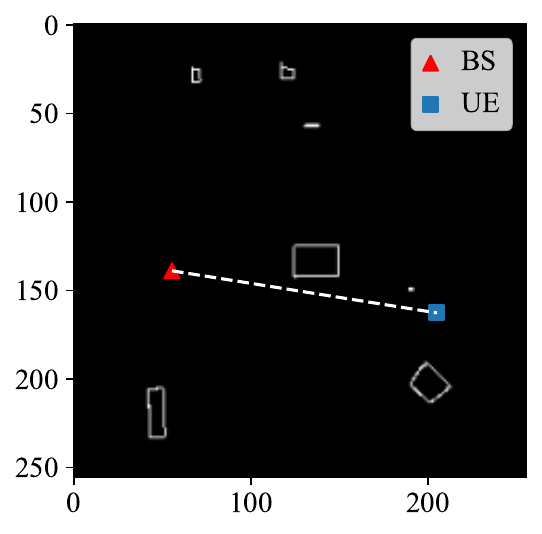}
		\end{minipage}}
	\caption{Left: Original pixel map, BS, and UE of area \#00955. Right: Pixel map after translating the obstacle building, BS, and UE. The similarities of comparing their features with the associated CSI feature are [0.564, 0.436].}
	\label{f7}
\end{figure}

First, we assess whether the pre-trained model can effectively discriminate between different BS-UE samples within an area. We compare the similarities of 150 environment-channel paired representations for 5 BSs and 30 UEs in each area of S1, as illustrated in Fig. \ref{f5}. The bright diagonal line indicates that most related modality pairs are correctly matched and distinguished from unrelated samples. Furthermore, we evaluate the environment-channel modality matching accuracy of the pre-trained model across 1,000 unseen S1 areas, as depicted in Fig. \ref{f5.5}. The x-axis represents matching accuracy intervals of 0.1, while the y-axis denotes the area counts within each interval. The statistical results indicate that our multi-modal universal model effectively differentiates samples in most unseen areas.

Next, we test the multi-user matching capability of the pre-trained model \cite{r10}. We compare the similarities between the representations of multiple UE positions and a given CSI. The BS and UE positions, along with the predicted probability distributions, are shown in Fig. \ref{f6}. Regardless of the number of UEs and their LOS or NLOS status, the likelihood of accurately identifying the correct position is significantly higher than that of identifying any incorrect positions.

Finally, we evaluate the environment sensing capability of the universal model. We compare the similarities between the representations of two environment data with different area maps and the UE CSI, as shown in Fig. \ref{f7}. The model effectively detects subtle changes in building positions. Additionally, we validate the impact of the scaling factor by comparing the features of three environment data with different physical-to-pixel scales \texttt{[}0.8s, s, 1.2s\texttt{]}, where the middle scale represents the true one, to the associated CSI features. Across all 250,000 unseen S1 and S2 samples, 90.11\% are correctly identified as matching the true scale among the three evaluated options. These results demonstrate that our pre-trained model can accurately infer the area physical size using only unseen normalized CSI, pixel map, and pixel positions.

Overall, our multi-modal universal model accurately captures the features of the physical environment and wireless channel data, as well as their underlying relationships.

\begin{table*}[t]
\caption{Task Types, Inputs, Outputs, and Adaptation Methods of the Four Exemplary Downstream Tasks}
\begin{center}
\begin{tabular}{ccccc}
\hline
\textbf{Downstream Task}&\textbf{Task Type}&\textbf{Input}&\textbf{Output}&\textbf{Adaptation Method}\\
\hline
CSI Feedback&Regression&CSI&CSI&ZSL\\
Direct Positioning&Regression&CSI&UE Position&TOFT\\
CSI/Position-based Beam Selection&Classification&CSI/Position&Beam Index&TOFT\\
LOS/NLOS Identification&Classification&CSI&UE LOS Status&ZSL, TOFT\\
\hline
\end{tabular}
\label{t2}
\end{center}
\end{table*}

\subsection{Downstream Tasks and Benchmarks}
Four typical use cases are discussed herein. CSI feedback enhancement aims to provide accurate CSI to BS with minimal uplink overhead, facilitating efficient beamforming \cite{r34}. High-precision direct positioning enhances services such as mobile navigation and autonomous driving \cite{r35}. CSI/position-based beam selection compensates for severe path loss caused by super-high frequencies, ensuring reliable services \cite{r36}. LOS/NLOS identification is critical for handover decisions and AI/non-AI methods selection for tasks such as positioning and beam management \cite{r37}. The task types, inputs, outputs, and adaptation methods for these tasks are provided in Table \ref{t2}.

The lightweight specific NNs for TOFT employ a multi-layer perceptron (MLP) architecture with interleaved fully connected layers and activation function layers. The choice of loss function for TOFT depends on the specific task. For the positioning task, the MSE loss is used and defined as follows,
\begin{equation}
L_{\mathrm{pos}}=\frac{1}{2N}\sum_{i=1}^{N}((\hat{x}_i-x_i)^2+(\hat{y}_i-y_i)^2),
\end{equation}
where $N$ denotes the number of samples, $x_i$ and $y_i$ represent the 2D coordinate labels, while $\hat{x}_i$ and $\hat{y}_i$ represent the predicted values for the $i$-th sample. For LOS/NLOS identification, cross-entropy loss is used and defined as follows,
\begin{equation}
L_{\mathrm{los}}=-\frac{1}{N}\sum_{i=1}^{N}(z_{i}\log{(p_{i})}+(1-z_{i})\log{(1-p_{i})}),
\end{equation}
where $z_{i}$ takes the label 0 for LOS and 1 for NLOS, while $p_{i}$ represents the probability of predicting NLOS for the $i$-th sample. To perform the beamforming task, we employ the widely adopted method of selecting the best beam index from a DFT codebook \cite{r38}. Due to the large number of beam indices, focal loss often outperforms cross-entropy loss \cite{r39}, so the loss function is formulated as,
\begin{equation}
L_{\mathrm{beam}} = -\frac{1}{N} \sum_{i=1}^{N} \sum_{j=1}^{N_t} m_{ij} (1 - q_{ij})^\gamma \log(q_{ij}),
\end{equation}
where $N_t$ denotes the number of transmitting antennas and also the number of beam indices, $m_{ij}$ is a binary indicator for whether the $j$-th beam index is the correct one for the $i$-th sample, $q_{ij}$ represents the probability of selecting the $j$-th index for the $i$-th sample, and $\gamma$ is the focusing parameter, which down-weights easy samples and focuses on hard samples.

To demonstrate the performance of the proposed paradigm across various downstream tasks in unseen areas and BSs, we consider the following benchmarks. For a fair comparison, the second benchmark is trained for 100 epochs using pre-training data to align with our pre-training model, while both our TOFT approach and the remaining benchmarks are trained for 1,000 epochs using 8,000 samples from each area unseen during pre-training. Our ZSL and TOFT methods and the five benchmarks are all tested on the same 2,000 remaining samples.
\begin{itemize}
\item Area-Specific ViT: This benchmark is used for CSI feedback and has the same model structure as CFENN and CRNN. It is trained using 8,000 single-area CSI samples, reflecting the extent of knowledge in the local area.
\item Area-General ViT: This benchmark is consistent with Area-Specific ViT except that it is trained using the pre-training CSI data outlined in Table \ref{t1}. It compresses and reconstructs CSIs from unseen areas directly, reflecting the richness of knowledge across extensive areas and the impact of our contrastive loss on reconstruction loss.
\item Task-Specific ViT: Separate models are adopted for the positioning, beam selection, and LOS/NLOS identification tasks. These models have structures and parameter sizes similar to CFENN, representing traditional area-, BS-, and task-specific methods.
\item Task-General ViT: Designed to perform three tasks simultaneously, this benchmark features a model structure and parameter size similar to CFENN but outputs UE position, LOS status, and beam index. The loss function is a combination of MSE, cross-entropy, and focal loss. This benchmark reflects the performance of scenario-specific but task-general methods. With similar model configurations, the performance of further generalization across areas and BSs would be inferior.
\item MLP: An MLP is used for position-based beam selection due to the low-dimensional position input and beam index output. Its latter part has the same structure as our lightweight fine-tuning NN, highlighting the benefits of knowledge injection through contrastive learning.
\end{itemize}

\subsection{Experimental Results and Analysis}

\begin{figure}[tb]
\centering
\includegraphics[width=1\columnwidth]{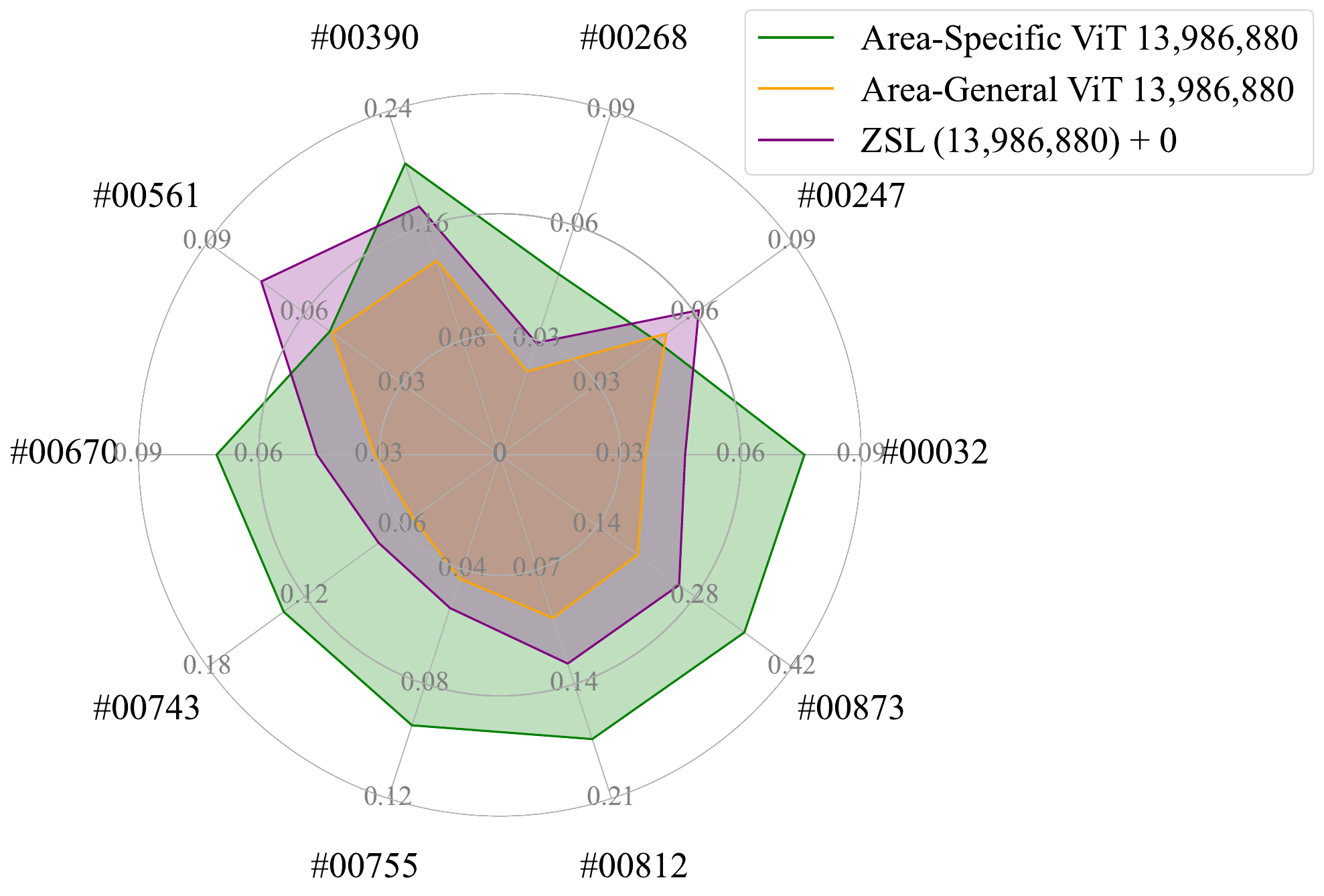}
\caption{The NMSE of CSI feedback and parameter size for Area-Specific ViT, Area-General ViT, and our ZSL methods in 10 unseen areas of S2.}
\label{f8}
\end{figure}

\textit{1) CSI Feedback:} The normalized MSE (NMSE) performance of the CSI feedback task and the number of parameters for the proposed scheme, Area-Specific ViT, and Area-General ViT in 10 unseen areas of S2 are shown in Fig. \ref{f8}. The NMSE of ZSL is significantly lower than that of the Area-Specific ViT and close to that of the Area-General ViT. The ZSL and Area-General ViT, with knowledge from other areas, achieve higher accuracy than the Area-Specific ViT with limited local knowledge, justifying the necessity of scenario generalization.

\begin{table*}[t]
\caption{Top-1 Accuracy and Parameter Size of the Beam Selection Task for Different Methods in 10 Unseen Areas of S2}
\begin{center}
\resizebox{\textwidth}{!}{
\begin{tabular}{ccccccccccc}
\hline
Method&\#00032&\#00247&\#00268&\#00390&\#00561&\#00670&\#00743&\#00755&\#00812&\#00873\\
\hline
Task-Specific ViT 7,125,024&95.55\%&94.70\%&96.55\%&94.45\%&97.15\%&95.85\%&95.40\%&95.65\%&94.35\%&93.25\%\\
Task-General ViT 7,126,052&86.75\%&87.65\%&85.50\%&74.30\%&87.55\%&75.20\%&77.25\%&68.40\%&67.40\%&70.70\%\\
\textbf{CSI-based TOFT} (7,117,312)+10,336&95.45\%&95.45\%&96.10\%&93.85\%&97.10\%&95.90\%&94.85\%&94.80\%&94.85\%&92.90\%\\
\hline
MLP 10,720&91.20\%&85.85\%&94.70\%&82.35\%&91.60\%&93.35\%&88.15\%&90.70\%&85.30\%&76.85\%\\
\textbf{Position-based TOFT} (7,518,720)+10,336&92.00\%&89.05\%&95.50\%&85.15\%&92.65\%&94.60\%&90.30\%&93.10\%&88.40\%&80.85\%\\
\hline
\end{tabular}}
\label{t3}
\end{center}
\end{table*}

\begin{figure}[tb]
\centering
\includegraphics[width=1\columnwidth]{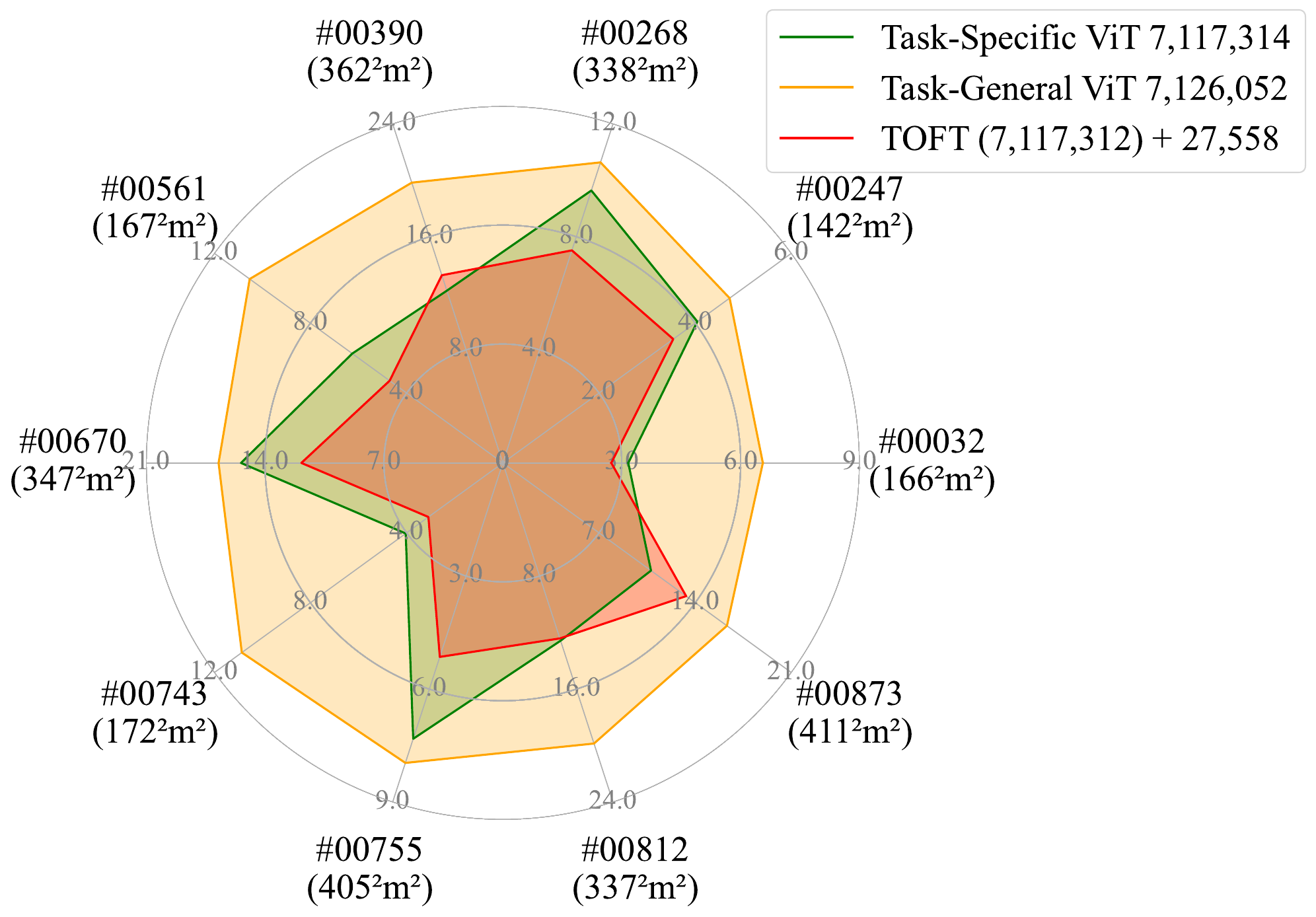}
\caption{The single-BS positioning error (at CDF90 in meters) and the number of parameters for Task-Specific ViT, Task-General ViT, and the proposed TOFT methods, as well as the area physical size, in 10 unseen areas of S2.}
\label{f9}
\end{figure}

\begin{figure}[t]
\centering
\includegraphics[width=1\columnwidth]{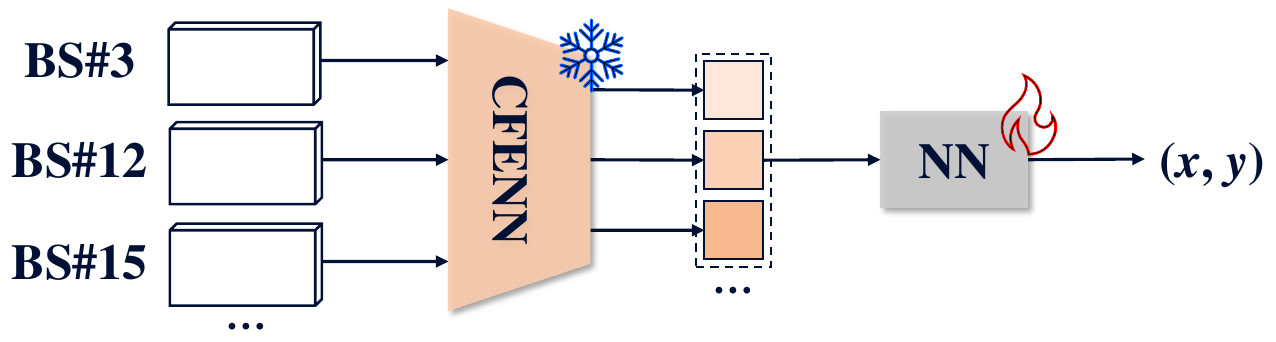}
\caption{The concatenation method of TOFT for multi-BS positioning.}
\label{f10}
\end{figure}

\begin{figure}[t]
\centering
\includegraphics[width=1\columnwidth]{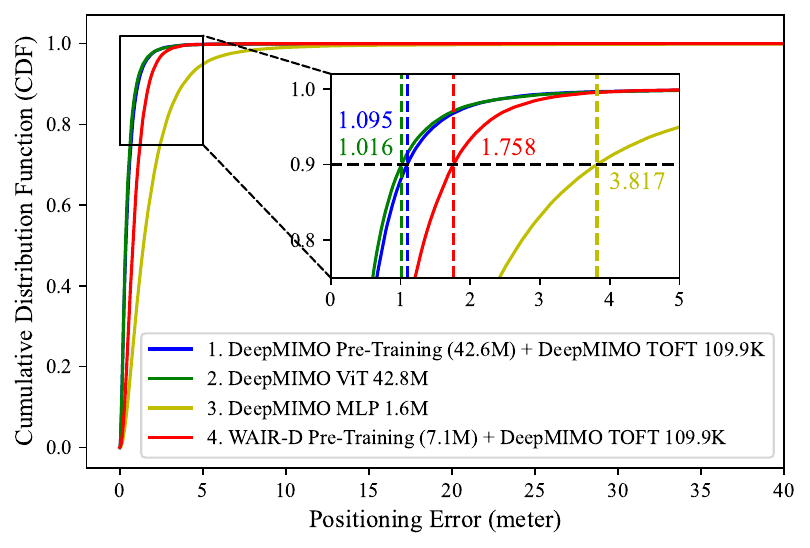}
\caption{The positioning error (at CDF, CDF90) and parameter size for the unprecedented 3-BS positioning task using the unseen DeepMIMO dataset.}
\label{f11}
\end{figure}

\begin{figure}[t]
\centering
\includegraphics[width=1\columnwidth]{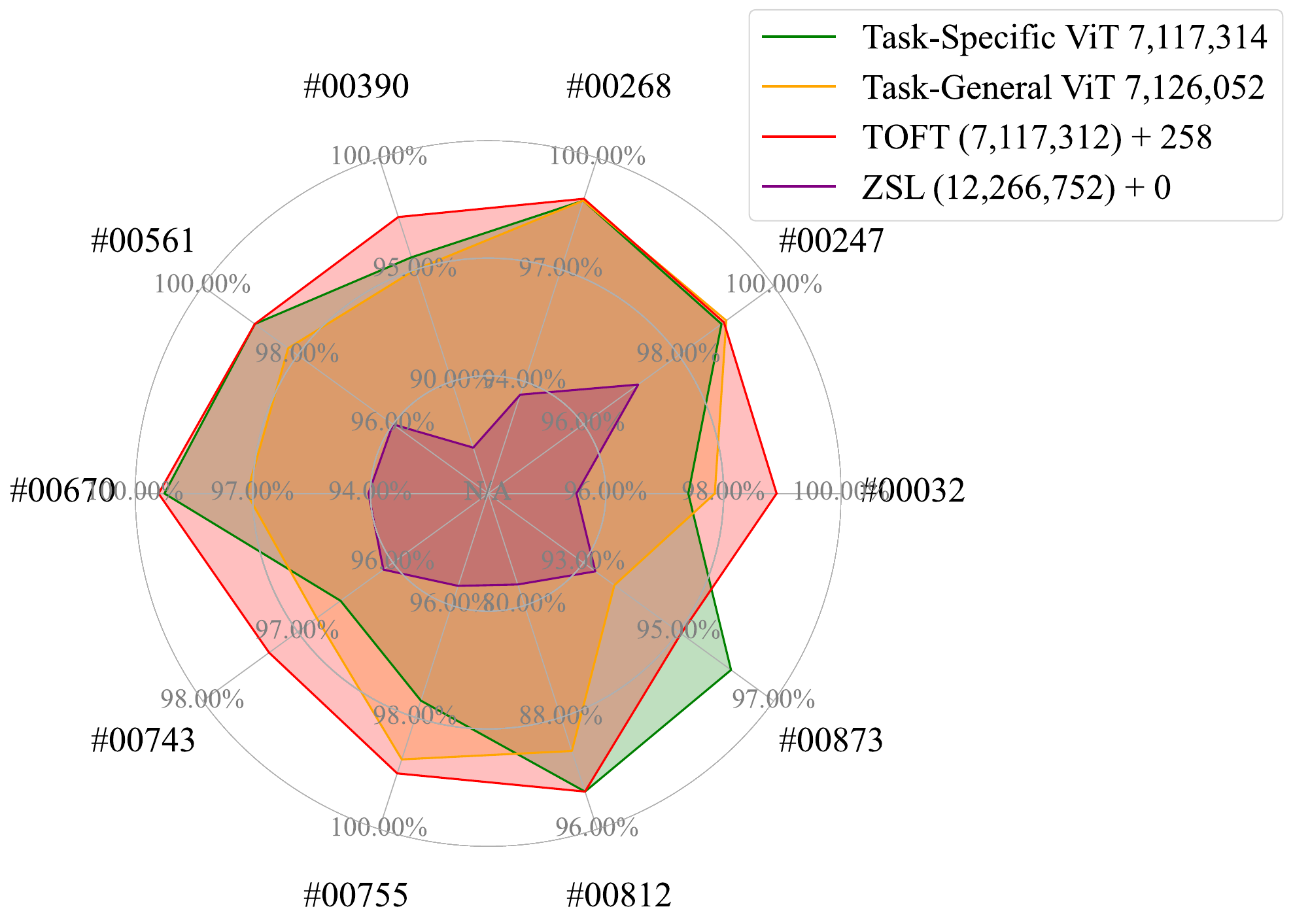}
\caption{The LOS/NLOS identification accuracy and parameter size for Task-Specific ViT, Task-General ViT, and our TOFT and ZSL in 10 unseen areas.}
\label{f12}
\end{figure}

\textit{2) Direct Positioning:} The positioning error corresponding to 90\% of the samples on the cumulative distribution function (CDF) curve of errors is denoted as CDF90. Fig. \ref{f9} depicts the single-BS positioning error and the number of parameters for Task-Specific ViT, Task-General ViT, and our TOFT methods, as well as the area physical size, in 10 unseen areas of S2. The proposed TOFT achieves an average positioning error at CDF90 that is 0.78m lower than that of Task-Specific ViT and 4.63m lower than that of Task-General ViT while using only 0.387\% of the tuning parameters of ViT. Task-General ViT performs worse than Task-Specific ViT because its model weights are affected by multiple task labels and loss functions.

In the following, we evaluate the generalization capability of the proposed framework in a cross-dataset and multi-BS scenario. Radio parameters such as carrier frequency, subcarrier number, and antenna configuration remain consistent between WAIR-D and the unseen DeepMIMO dataset \cite{r40}. The DeepMIMO Outdoor 1 (O1) scenario is employed, which includes 18 BSs. The UEs are located in a cross-shaped area surrounded by buildings of varying sizes. Positioning error can be reduced using CSIs from multiple BSs. However, current methods cannot accommodate situations with varying numbers of BSs and flexible BS combinations due to their BS-specific end-to-end training strategy. Although our paradigm aligns CSI with a BS-UE position, it offers scalability to multi-BS scenarios. As depicted in Fig. \ref{f10}, CSIs from different unseen BSs are separately fed into the frozen CFENN. All output representations are concatenated and fed into a lightweight NN for TOFT. The 3-BS positioning error, particularly at CDF90, and the number of parameters for different methods are presented in Fig. \ref{f11}. The first method involves earlier pre-training with 3.8M samples from 15 BSs in DeepMIMO, followed by TOFT with 167,220 samples from unseen BS\#3, BS\#12, and BS\#15. The second method uses a ViT trained from scratch, similar in structure and parameter size to the pre-training model of the first method. The third method uses a simple, low-cost MLP model. The fourth method applies CFENN pre-trained on WAIR-D to process DeepMIMO data, followed by TOFT with the same fine-tuning NN structure as in the first method. The positioning error at CDF90 for our fourth method is 1.758m, close to the 1.095m and 1.016m achieved by the first and second methods, and significantly better than the 3.817m of the third method, while requiring only 0.257\% of the tuning parameters of ViT. Our universal model is applicable to unseen and spanning datasets and is more flexible and cost-effective.

\textit{3) CSI/Position-Based Beam Selection:} The top-1 beam selection accuracy and parameter size for our CSI/position-based TOFT methods and three benchmarks in 10 unseen areas are shown in Table \ref{t3}. CSI-based TOFT achieves similar accuracy to Task-Specific ViT and significantly outperforms Task-General ViT while using only 0.145\% of ViT's tuning parameters. Leveraging knowledge from EPNN, position-based TOFT consistently surpasses MLP in accuracy. In areas with sparse buildings, UE positions sharing the same optimal beam index form a BS-centered sector-shaped cluster \cite{r30}, which can be easily modeled by MLP, thereby achieving high accuracy. However, in more complex physical environments, MLP's accuracy decreases, while the accuracy of the proposed position-based TOFT scheme significantly improves.


\textit{4) LOS/NLOS Identification:} To demonstrate that our universal model can accurately determine UEs' LOS status across thousands of unseen areas and BSs, we first evaluate our ZSL method using all 250,000 unseen samples from the ZSL and TOFT dataset (Table \ref{t1}) without any tuning parameters. ZSL achieves an impressive average classification accuracy of 92.31\%. Then, Fig. \ref{f12} presents the classification accuracy and parameter size for Task-Specific ViT, Task-General ViT, and our TOFT and ZSL methods in 10 unseen areas of S2. ZSL achieves accuracy exceeding 90\% in most areas. TOFT outperforms both Task-Specific ViT and Task-General ViT while using only 0.0036\% of ViT's tuning parameters. Moreover, fine-tuning 258 parameters results in faster convergence.

The superior performance of TOFT in different task objectives proves the compatibility of universal representations.

\section{Conclusion}
In this paper, we propose a transformative multi-modal pre-training and downstream task adaptation paradigm for flexibly and cost-effectively performing multiple wireless communication tasks across diverse scenarios. In the pre-training stage, multiple related modalities, such as the physical environment and wireless channel, are effectively aligned to achieve cross-modality knowledge fusion and extract universal representations. Both qualitative and quantitative analyses are conducted to validate the effectiveness of the proposed contrastive method in multi-modality interaction. In the downstream task adaptation stage, the direct ZSL and pluggable TOFT methods are devised to flexibly perform various tasks with extremely low costs. Experimental results corroborate that our paradigm outperforms conventional task-specific and multi-area/output methods for exemplary sub-tasks in various unseen scenarios while utilizing fewer than 0.387\% of their tuning parameters.

Our proposed paradigm effectively integrates diverse modalities in future AI-native wireless systems, heralding a promising future for unified communication and sensing intelligence.


\vfill
\end{document}